\documentclass[twocolumn]{aastex631}

\shorttitle{HD~136164~Ab GRAVITY}
\shortauthors{Balmer et al.}
\graphicspath{{./}{figures/}}
\definecolor{mylinkcolor}{RGB}{16, 37, 110} % my color
\hypersetup{linkcolor=mylinkcolor,citecolor=mylinkcolor,urlcolor=mylinkcolor}

\received{Oct. 11th, 2023}
\revised{Nov. 29th, 2023}
\accepted{Dec. 13th, 2023}

\submitjournal{The Astronomical Journal}

\begin{document}

% \title{Interferometric View of the Young, Substellar Companion HD~136164~Ab}
\title{VLTI/GRAVITY Provides Evidence the Young, Substellar Companion HD~136164~Ab formed like a ``Failed Star"}

\author[0000-0001-6396-8439]{William O. Balmer}
\correspondingauthor{William O. Balmer}
\email{wbalmer1@jhu.edu}
\altaffiliation{Johns Hopkins University George Owen Fellow}
\affiliation{Department of Physics \& Astronomy, Johns Hopkins University, 3400 N. Charles Street, Baltimore, MD 21218, USA}
\affiliation{ Space Telescope Science Institute, 3700 San Martin Drive, Baltimore, MD 21218, USA}
\author{L.~Pueyo}
\affiliation{ Space Telescope Science Institute, 3700 San Martin Drive, Baltimore, MD 21218, USA}
\author{S.~Lacour}
\affiliation{ LESIA, Observatoire de Paris, PSL, CNRS, Sorbonne Universit\'e, Universit\'e de Paris, 5 place Janssen, 92195 Meudon, France}
\affiliation{ European Southern Observatory, Karl-Schwarzschild-Stra\ss e 2, 85748 Garching, Germany}
\author{J.~J.~Wang}
\affiliation{ Center for Interdisciplinary Exploration and Research in Astrophysics (CIERA) and Department of Physics and Astronomy, Northwestern University, Evanston, IL 60208, USA}
\author{T.~Stolker}
\affiliation{ Leiden Observatory, Leiden University, P.O. Box 9513, 2300 RA Leiden, The Netherlands}
\author{J.~Kammerer}
\affiliation{ Space Telescope Science Institute, 3700 San Martin Drive, Baltimore, MD 21218, USA}
\author{N.~Pourr\'e}
\affiliation{ Universit\'e Grenoble Alpes, CNRS, IPAG, 38000 Grenoble, France}
\author{M.~Nowak}
\affiliation{ Institute of Astronomy, University of Cambridge, Madingley Road, Cambridge CB3 0HA, United Kingdom}
\author{E.~Rickman}
\affiliation{ European Space Agency (ESA), ESA Office, Space Telescope Science Institute, 3700 San Martin Drive, Baltimore, MD 21218, USA}
\author{S.~Blunt}
\affiliation{ Center for Interdisciplinary Exploration and Research in Astrophysics (CIERA) and Department of Physics and Astronomy, Northwestern University, Evanston, IL 60208, USA}
\author{A.~Sivaramakrishnan}
\affiliation{ Space Telescope Science Institute, 3700 San Martin Drive, Baltimore, MD 21218, USA}
\affiliation{ Department of Physics \& Astronomy, Johns Hopkins University, 3400 N. Charles Street, Baltimore, MD 21218, USA}
\author{D.~Sing}
\affiliation{ Department of Physics \& Astronomy, Johns Hopkins University, 3400 N. Charles Street, Baltimore, MD 21218, USA}
\affiliation{ Department of Earth \& Planetary Sciences, Johns Hopkins University, Baltimore, MD, USA}
\author{K.~Wagner}
\affiliation{ Department of Astronomy and Steward Observatory, University of Arizona, 933 N Cherry Ave, Tucson, AZ 85712. USA}
\author{G.-D.~Marleau}
\affiliation{ Fakult\"at f\"ur Physik, Universit\"at Duisburg-Essen, Lotharstraße 1, 47057 Duisburg, Germany}
\affiliation{ Instit\"ut f\"ur Astronomie und Astrophysik, Universit\"at T\"ubingen, Auf der Morgenstelle 10, 72076 T\"ubingen, Germany}
\affiliation{ Center for Space and Habitability, Universit\"at Bern, Gesellschaftsstrasse 6, 3012 Bern, Switzerland}
\affiliation{ Max Planck Institute for Astronomy, K\"onigstuhl 17, 69117 Heidelberg, Germany}
\author{A.-M.~Lagrange}
\affiliation{ Universit\'e Grenoble Alpes, CNRS, IPAG, 38000 Grenoble, France}
\affiliation{ LESIA, Observatoire de Paris, PSL, CNRS, Sorbonne Universit\'e, Universit\'e de Paris, 5 place Janssen, 92195 Meudon, France}
\author{R.~Abuter}
\affiliation{ European Southern Observatory, Karl-Schwarzschild-Stra\ss e 2, 85748 Garching, Germany}
\author{A.~Amorim}
\affiliation{ Universidade de Lisboa - Faculdade de Ci\^encias, Campo Grande, 1749-016 Lisboa, Portugal}
\affiliation{ CENTRA - Centro de Astrof\' isica e Gravita\c c\~ao, IST, Universidade de Lisboa, 1049-001 Lisboa, Portugal}
\author{R.~Asensio-Torres}
\affiliation{ Max Planck Institute for Astronomy, K\"onigstuhl 17, 69117 Heidelberg, Germany}
\author{J.-P.~Berger}
\affiliation{ Universit\'e Grenoble Alpes, CNRS, IPAG, 38000 Grenoble, France}
\author{H.~Beust}
\affiliation{ Universit\'e Grenoble Alpes, CNRS, IPAG, 38000 Grenoble, France}
\author{A.~Boccaletti}
\affiliation{ LESIA, Observatoire de Paris, PSL, CNRS, Sorbonne Universit\'e, Universit\'e de Paris, 5 place Janssen, 92195 Meudon, France}
\author{A.~Bohn}
\affiliation{ Leiden Observatory, Leiden University, P.O. Box 9513, 2300 RA Leiden, The Netherlands}
\author{M.~Bonnefoy}
\affiliation{ Universit\'e Grenoble Alpes, CNRS, IPAG, 38000 Grenoble, France}
\author{H.~Bonnet}
\affiliation{ European Southern Observatory, Karl-Schwarzschild-Stra\ss e 2, 85748 Garching, Germany}
\author{M.~S.~Bordoni}
\affiliation{ Max Planck Institute for extraterrestrial Physics, Giessenbachstra\ss e~1, 85748 Garching, Germany}
\author{G.~Bourdarot}
\affiliation{ Max Planck Institute for extraterrestrial Physics, Giessenbachstra\ss e~1, 85748 Garching, Germany}
\author{W.~Brandner}
\affiliation{ Max Planck Institute for Astronomy, K\"onigstuhl 17, 69117 Heidelberg, Germany}
\author{F.~Cantalloube}
\affiliation{ Aix Marseille Univ, CNRS, CNES, LAM, Marseille, France}
\author{P.~Caselli }
\affiliation{ Max Planck Institute for extraterrestrial Physics, Giessenbachstra\ss e~1, 85748 Garching, Germany}
\author{B.~Charnay}
\affiliation{ LESIA, Observatoire de Paris, PSL, CNRS, Sorbonne Universit\'e, Universit\'e de Paris, 5 place Janssen, 92195 Meudon, France}
\author{G.~Chauvin}
\affiliation{ Universit\'e Grenoble Alpes, CNRS, IPAG, 38000 Grenoble, France}
\author{A.~Chavez}
\affiliation{ Center for Interdisciplinary Exploration and Research in Astrophysics (CIERA) and Department of Physics and Astronomy, Northwestern University, Evanston, IL 60208, USA}
\author{E.~Choquet}
\affiliation{ Aix Marseille Univ, CNRS, CNES, LAM, Marseille, France}
\author{V.~Christiaens}
\affiliation{ STAR Institute/Universit\'e de Li\`ege, Belgium}
\author{Y.~Cl\'enet}
\affiliation{ LESIA, Observatoire de Paris, PSL, CNRS, Sorbonne Universit\'e, Universit\'e de Paris, 5 place Janssen, 92195 Meudon, France}
\author{V.~Coud\'e~du~Foresto}
\affiliation{ LESIA, Observatoire de Paris, PSL, CNRS, Sorbonne Universit\'e, Universit\'e de Paris, 5 place Janssen, 92195 Meudon, France}
\author{A.~Cridland}
\affiliation{ Leiden Observatory, Leiden University, P.O. Box 9513, 2300 RA Leiden, The Netherlands}
\author{R.~Davies}
\affiliation{ Max Planck Institute for extraterrestrial Physics, Giessenbachstra\ss e~1, 85748 Garching, Germany}
\author{R.~Dembet}
\affiliation{ LESIA, Observatoire de Paris, PSL, CNRS, Sorbonne Universit\'e, Universit\'e de Paris, 5 place Janssen, 92195 Meudon, France}
\author{A.~Drescher}
\affiliation{ Max Planck Institute for extraterrestrial Physics, Giessenbachstra\ss e~1, 85748 Garching, Germany}
\author{G.~Duvert}
\affiliation{ Universit\'e Grenoble Alpes, CNRS, IPAG, 38000 Grenoble, France}
\author{A.~Eckart}
\affiliation{ 1.\ Institute of Physics, University of Cologne, Z\"ulpicher Stra\ss e 77, 50937 Cologne, Germany}
\affiliation{ Max Planck Institute for Radio Astronomy, Auf dem H\"ugel 69, 53121 Bonn, Germany}
\author{F.~Eisenhauer}
\affiliation{ Max Planck Institute for extraterrestrial Physics, Giessenbachstra\ss e~1, 85748 Garching, Germany}
\author{N.~M.~F"orster Schreiber}
\affiliation{ Max Planck Institute for extraterrestrial Physics, Giessenbachstra\ss e~1, 85748 Garching, Germany}
\author{P.~Garcia}
\affiliation{ CENTRA - Centro de Astrof\' isica e Gravita\c c\~ao, IST, Universidade de Lisboa, 1049-001 Lisboa, Portugal}
\affiliation{ Universidade do Porto, Faculdade de Engenharia, Rua Dr.~RobertoRua Dr.~Roberto Frias, 4200-465 Porto, Portugal}
\author{R.~Garcia~Lopez}
\affiliation{ School of Physics, University College Dublin, Belfield, Dublin 4, Ireland}
\affiliation{ Max Planck Institute for Astronomy, K\"onigstuhl 17, 69117 Heidelberg, Germany}
\author{E.~Gendron}
\affiliation{ LESIA, Observatoire de Paris, PSL, CNRS, Sorbonne Universit\'e, Universit\'e de Paris, 5 place Janssen, 92195 Meudon, France}
\author{R.~Genzel}
\affiliation{ Max Planck Institute for extraterrestrial Physics, Giessenbachstra\ss e~1, 85748 Garching, Germany}
\author{S.~Gillessen}
\affiliation{ Max Planck Institute for extraterrestrial Physics, Giessenbachstra\ss e~1, 85748 Garching, Germany}
\author{J.~H.~Girard}
\affiliation{ Space Telescope Science Institute, 3700 San Martin Drive, Baltimore, MD 21218, USA}
\author{S.~Grant}
\affiliation{ Max Planck Institute for extraterrestrial Physics, Giessenbachstra\ss e~1, 85748 Garching, Germany}
\author{X.~Haubois}
\affiliation{ European Southern Observatory, Casilla 19001, Santiago 19, Chile}
\author{G.~Hei\ss el}
\affiliation{ Advanced Concepts Team, European Space Agency, TEC-SF, ESTEC, Keplerlaan 1, NL-2201, AZ Noordwijk, The Netherlands}
\affiliation{ LESIA, Observatoire de Paris, PSL, CNRS, Sorbonne Universit\'e, Universit\'e de Paris, 5 place Janssen, 92195 Meudon, France}
\author{Th.~Henning}
\affiliation{ Max Planck Institute for Astronomy, K\"onigstuhl 17, 69117 Heidelberg, Germany}
\author{S.~Hinkley}
\affiliation{ University of Exeter, Physics Building, Stocker Road, Exeter EX4 4QL, United Kingdom}
\author{S.~Hippler}
\affiliation{ Max Planck Institute for Astronomy, K\"onigstuhl 17, 69117 Heidelberg, Germany}
\author{M.~Houll\'e}
\affiliation{ Université Côte d’Azur, Observatoire de la Côte d’Azur, CNRS, Laboratoire Lagrange, France}
\author{Z.~Hubert}
\affiliation{ Universit\'e Grenoble Alpes, CNRS, IPAG, 38000 Grenoble, France}
\author{L.~Jocou}
\affiliation{ Universit\'e Grenoble Alpes, CNRS, IPAG, 38000 Grenoble, France}
\author{M.~Keppler}
\affiliation{ Max Planck Institute for Astronomy, K\"onigstuhl 17, 69117 Heidelberg, Germany}
\author{P.~Kervella}
\affiliation{ LESIA, Observatoire de Paris, PSL, CNRS, Sorbonne Universit\'e, Universit\'e de Paris, 5 place Janssen, 92195 Meudon, France}
\author{L.~Kreidberg}
\affiliation{ Max Planck Institute for Astronomy, K\"onigstuhl 17, 69117 Heidelberg, Germany}
\author{N.~T.~Kurtovic}
\affiliation{ Max Planck Institute for extraterrestrial Physics, Giessenbachstra\ss e~1, 85748 Garching, Germany}
\author{V.~Lapeyr\`ere}
\affiliation{ LESIA, Observatoire de Paris, PSL, CNRS, Sorbonne Universit\'e, Universit\'e de Paris, 5 place Janssen, 92195 Meudon, France}
\author{J.-B.~Le~Bouquin}
\affiliation{ Universit\'e Grenoble Alpes, CNRS, IPAG, 38000 Grenoble, France}
\author{P.~L\'ena}
\affiliation{ LESIA, Observatoire de Paris, PSL, CNRS, Sorbonne Universit\'e, Universit\'e de Paris, 5 place Janssen, 92195 Meudon, France}
\author{D.~Lutz}
\affiliation{ Max Planck Institute for extraterrestrial Physics, Giessenbachstra\ss e~1, 85748 Garching, Germany}
\author{A.-L.~Maire}
\affiliation{ Universit\'e Grenoble Alpes, CNRS, IPAG, 38000 Grenoble, France}
\author{F.~Mang}
\affiliation{ Max Planck Institute for extraterrestrial Physics, Giessenbachstra\ss e~1, 85748 Garching, Germany}
\author{A.~M\'erand}
\affiliation{ European Southern Observatory, Karl-Schwarzschild-Stra\ss e 2, 85748 Garching, Germany}
\author{P.~Molli\`ere}
\affiliation{ Max Planck Institute for Astronomy, K\"onigstuhl 17, 69117 Heidelberg, Germany}
\author{C.~Mordasini}
\affiliation{ Center for Space and Habitability, Universit"at Bern, Gesellschaftsstrasse 6, 3012 Bern, Switzerland}
\author{D.~Mouillet}
\affiliation{ Universit\'e Grenoble Alpes, CNRS, IPAG, 38000 Grenoble, France}
\author{E.~Nasedkin}
\affiliation{ Max Planck Institute for Astronomy, K\"onigstuhl 17, 69117 Heidelberg, Germany}
\author{T.~Ott}
\affiliation{ Max Planck Institute for extraterrestrial Physics, Giessenbachstra\ss e~1, 85748 Garching, Germany}
\author{G.~P.~P.~L.~Otten}
\affiliation{ Academia Sinica, Institute of Astronomy and Astrophysics, 11F Astronomy-Mathematics Building, NTU/AS campus, No. 1, Section 4, Roosevelt Rd., Taipei 10617, Taiwan}
\author{C.~Paladini}
\affiliation{ European Southern Observatory, Casilla 19001, Santiago 19, Chile}
\author{T.~Paumard}
\affiliation{ LESIA, Observatoire de Paris, PSL, CNRS, Sorbonne Universit\'e, Universit\'e de Paris, 5 place Janssen, 92195 Meudon, France}
\author{K.~Perraut}
\affiliation{ Universit\'e Grenoble Alpes, CNRS, IPAG, 38000 Grenoble, France}
\author{G.~Perrin}
\affiliation{ LESIA, Observatoire de Paris, PSL, CNRS, Sorbonne Universit\'e, Universit\'e de Paris, 5 place Janssen, 92195 Meudon, France}
\author{O.~Pfuhl}
\affiliation{ European Southern Observatory, Karl-Schwarzschild-Stra\ss e 2, 85748 Garching, Germany}
\author{D.~C.~Ribeiro}
\affiliation{ Max Planck Institute for extraterrestrial Physics, Giessenbachstra\ss e~1, 85748 Garching, Germany}
\author{L.~Rodet}
\affiliation{ Center for Astrophysics and Planetary Science, Department of Astronomy, Cornell University, Ithaca, NY 14853, USA}
\author{Z.~Rustamkulov}
\affiliation{ Department of Earth \& Planetary Sciences, Johns Hopkins University, Baltimore, MD, USA}
\author{J.~Shangguan}
\affiliation{ Max Planck Institute for extraterrestrial Physics, Giessenbachstra\ss e~1, 85748 Garching, Germany}
\author{T.~Shimizu }
\affiliation{ Max Planck Institute for extraterrestrial Physics, Giessenbachstra\ss e~1, 85748 Garching, Germany}
\author{C.~Straubmeier}
\affiliation{ 1.\ Institute of Physics, University of Cologne, Z\"ulpicher Stra\ss e 77, 50937 Cologne, Germany}
\author{E.~Sturm}
\affiliation{ Max Planck Institute for extraterrestrial Physics, Giessenbachstra\ss e~1, 85748 Garching, Germany}
\author{L.~J.~Tacconi}
\affiliation{ Max Planck Institute for extraterrestrial Physics, Giessenbachstra\ss e~1, 85748 Garching, Germany}
\author{A.~Vigan}
\affiliation{ Aix Marseille Univ, CNRS, CNES, LAM, Marseille, France}
\author{F.~Vincent}
\affiliation{ LESIA, Observatoire de Paris, PSL, CNRS, Sorbonne Universit\'e, Universit\'e de Paris, 5 place Janssen, 92195 Meudon, France}
\author{K.~Ward-Duong}
\affiliation{ Department of Astronomy, Smith College, Northampton MA 01063 USA}
\author{F.~Widmann}
\affiliation{ Max Planck Institute for extraterrestrial Physics, Giessenbachstra\ss e~1, 85748 Garching, Germany}
\author{T.~Winterhalder}
\affiliation{ European Southern Observatory, Karl-Schwarzschild-Stra\ss e 2, 85748 Garching, Germany}
\author{J.~Woillez}
\affiliation{ European Southern Observatory, Karl-Schwarzschild-Stra\ss e 2, 85748 Garching, Germany}
\author{S.~Yazici}
\affiliation{ Max Planck Institute for extraterrestrial Physics, Giessenbachstra\ss e~1, 85748 Garching, Germany}
\author{the GRAVITY Collaboration}

\begin{abstract}

Young, low-mass Brown Dwarfs orbiting early-type stars, with low mass ratios ($q\lesssim0.01$), appear intrinsically rare and present a formation dilemma: could a handful of these objects be the highest mass outcomes of ``planetary" formation channels (bottom up within a protoplanetary disk), or are they more representative of the lowest mass ``failed binaries" (formed via disk fragmentation, or core fragmentation)? Additionally, their orbits can yield model-independent dynamical masses, and when paired with wide wavelength coverage and accurate system age estimates, can constrain evolutionary models in a regime where the models have a wide dispersion depending on initial conditions. We present new interferometric observations of the $16\,\mathrm{Myr}$ substellar companion HD~136164~Ab (HIP~75056~Ab) with VLTI/GRAVITY and an updated orbit fit including proper motion measurements from the \textit{Hipparcos}-\textit{Gaia} Catalogue of Accelerations. We estimate a dynamical mass of $35\pm10\,\mathrm{M_J}$ ($q\sim0.02$), making HD~136164~Ab the youngest substellar companion with a dynamical mass estimate. The new mass and newly constrained orbital eccentricity ($e=0.44\pm0.03$) and separation ($22.5\pm1\,\mathrm{au}$) could indicate that the companion formed via the low-mass tail of the Initial Mass Function. Our atmospheric fit to the \texttt{SPHINX} M-dwarf model grid suggests a sub-solar C/O ratio of $0.45$, and $3\times$ solar metallicity, which could indicate formation in the circumstellar disk via disk fragmentation. Either way, the revised mass estimate likely excludes ``bottom-up" formation via core accretion in the circumstellar disk. HD~136164~Ab joins a select group of young substellar objects with dynamical mass estimates; epoch astrometry from future \textit{Gaia} data releases will constrain the dynamical mass of this crucial object further.

\end{abstract}

% \keywords{Classical Novae (251) --- Ultraviolet astronomy(1736) --- History of astronomy(1868) --- Interdisciplinary astronomy(804)}


\section{Introduction} \label{sec:intro}

\par The origins of low mass brown dwarf companions to stars are uncertain. These objects could form from the low mass tail of the Initial Mass Function \citep[IMF, see e.g.][]{Chabrier2003}, via fragmentation of a molecular cloud, and be captured into binary orbits \citep[e.g.][]{Padoan2004, Boyd2005, Bate2009, Bate2012}, or they could form via the gravitational instability of a circumstellar disk \citep{Boss1997, Stamatellos2007, Stamatellos2009, Kratter2010, Forgan2013, Forgan2015, Forgan2018}. Some even argue that a handful of these objects might form via core accretion like planets, and undergo runaway accretion \citep{D'Angelo2010, Molliere2012, Bodenheimer2013} to reach masses greater than the deuterium burning limit \citep[$\simeq13~\mathrm{M_J}$,][]{Saumon1996, Chabrier2000}. \citet{Schlaufman2018} argues to the contrary, showing that planetary mass companions with masses below $4~\mathrm{M_J}$ orbit stars with on average higher metallicity (associated with a larger amount of material for core accretion formation), while planetary mass companions with masses above this cutoff do not necessarily orbit metal-rich stars. This would appear to indicate there is population-level evidence for distinct formation channels for exoplanets and brown dwarf companions with masses above and below $4~\mathrm{M_J}$. 
\par There are a number of curious systems that blur these observational rules-of-thumb that enforce the boundaries between formation channels, for instance the HD~206893 system, where two nearly co-planar low mass companions reside within a debris disk \citep{Milli2017, Delorme2017, Grandjean2019, Ward-Duong2021, Kammerer2021, Hinkley2023}. For newly discovered low mass companions, especially low mass companions to early type stars (who have low mass ratios), this confusion is inevitable. However, further orbital characterization can provide clues to their formation \citep{Bowler2020}, and spectral characterization can provide a further line of evidence. 
\par Additionally, evolutionary and spectral models of substellar objects (brown dwarfs, massive giant planets), like their higher mass stellar counterparts \citep[e.g.][]{Rodet2018, Dieterich2021}, need to be tested and refined based on model independent measurements of their masses derived from orbital characterization \citep{Dupuy2017, Brandt2019, Fontanive2019, Rickman2020, Vrijmoet2020, mBrandt2021, Rickman2022, Bonavita2022, Franson2022, Vrijmoet2022, Franson2023, Balmer2023}. The difficulty is that substellar objects amenable to these measurements, unlike stellar binaries, appear intrinsically rare \citep[e.g.][]{Dieterich2012, Fontanive2018, Nielsen2019, Duchene2023}, and generally substellar companions are orders of magnitude fainter than their hosts, necessitating a greater observational expense than stellar binary orbits. So far, the number of substellar objects with dynamically derived masses is low, about 20 in total, and the number of these with accurate age estimates younger than a few hundred megayears is of order unity \citep[see Figure 4 in][]{Franson2023}. The youngest brown dwarf with a dynamical mass, PZ Tel b \citep[$27^{+25}_{-9}\,\mathrm{M_J}$, $24\pm3\,\mathrm{Myr}$][]{Franson2023}, has been predicted by evolutionary models to have a higher mass on average than its dynamical mass. Older, cold brown dwarfs like HD~19467~B have been found to be under-luminous compared to evolutionary models \citep{mBrandt2021, Greenbaum2023}. It remains to be seen to what extent there are systematic discrepancies between dynamical and evolutionary models with age, as more dynamical masses need to be estimated for each age category, particularly for young ages.

\par The sensitivity of direct imaging to higher intrinsic luminosities has revealed many faint, substellar companions to stars in young stellar associations, like the greater Scorpius-Centarus (Sco-Cen) association \citep{Janson2012, Hinkley2015, Cheetham2018, Bohn2022}. These discoveries present an excellent opportunity to determine and study the dynamical masses of young brown dwarfs. 

\par HD~136164~A (HIP~75056~A, 2MASS~J15201339-3455316, GAIA~DR3~6206053714943873408) is a bright \citep[$G_{mag}=7.75$,][]{GaiaCollaboration2022}, A2V type \citep{Houk1982} member of the Upper-Centaurus-Lupus (UCL) association \citep[with 99.9\% probability membership according to BANYAN $\Sigma$,][]{Gagne2018}. HD~136164~A has a \textit{Gaia} DR3 parallax of $8.20\pm0.04$, corresponding to a distance of $122~\mathrm{pc}$. \citet{Pecaut2016} report an age for UCL of $16\pm2~\mathrm{Myr}$ based on a multiwavelength spectroscopic and photometric evolutionary study of the association; \citet{Zerjal2023} report an age of $15\pm3~\mathrm{Myr}$, based purely on kinematics. For the primary mass, we adopt $M=1.7\pm0.1\,\mathrm{M_\odot}$ from \citet{Kervella2022}, based on \citet{Girardi2000}.

\par The system has a low \textit{Gaia} RUWE (1.053), but has a strong proper-motion anomaly between \textit{\textit{Hipparcos}} and \textit{Gaia} noted in \citet{Kervella2019, Kervella2022} and a $\chi^2=15.9$ in the \textit{Hipparcos}-\textit{Gaia} Catalog of Accelerations \citep[HGCA,][]{Brandt2021}, indicative of the presence of a perturbing companion. \citet{Kouwenhoven2007} and \citet{Wagner2020} identify a nearby low mass star (HD~136164~AB, $\mathrm{M_B}\sim0.3~\mathrm{M_\odot}$, $\mathrm{Ks_{mag}}=~11.2$) at a separation of 5\farcs2 ($\sim650~\mathrm{au}$), which cannot reproduce the observed astrometric signal \citep{Kervella2022}.

\par \citet{Wagner2020} first identified the substellar companion HD~136164~Ab as part of their coronagraphic survey of A-type stars in Sco-Cen \citep{Wagner2022}. At a separation of $125-150~\mathrm{mas}$, the companion is consistent with inducing the astrometric signal observed between \textit{\textit{Hipparcos}} and \textit{Gaia}; at a separation of $\sim30~\mathrm{au}$ the acceleration is consistent with a substellar mass \citep{Kervella2022}. The NIR magnitudes are consistent with the companion being substellar \citep[give $25\pm5~\mathrm{M_J}$ based on evolutionary models from \citet{Baraffe2003}]{Wagner2020}, and \citet{Wagner2020} report an estimate of the companion's spectral type (M6-L2) consistent with the observed absorption due to water in the low resolution SPHERE spectrophotometry. Only considering two epochs of astrometry, they partially constrain the separation of the companion's orbit ($\sim30\pm15~\mathrm{au}$), and were unable to estimate the eccentricity reliably. 
\par With a low mass ratio and close separation, \citet{Wagner2020} posit that the companion could have formed via disk fragmentation \citep[e.g.][]{Boss1997, Kratter2010, Forgan2018}, or even the very high end of a ``planetary" core-accretion formation channel \citep[e.g.][]{Pollack1996, Mordasini2012} although the possibility \citep[e.g.][]{Emsenhuber2021} and occurrence \citep[e.g.][]{Schlaufman2018} of such high mass core-accretion planets is still heavily debated. Placing the companion's orbit in context could provide insight into this question, as recent population level studies of substellar companions indicate two populations in eccentricity \citep{Bowler2020, Nagpal2023, DoO2023}. It appears that low mass objects (that likely formed within a protoplanetary disk, whose gas damped their initial eccentricity) exhibit a distribution of eccentricities that is on average lower than the distribution for higher mass objects (that likely formed via core fragmentation along with the stellar population, and formed within or became captured into binary orbits). Coupling a measurement of the mass and eccentricity of the object from a Keplerian orbit could therefore inform the interpretation of its formation history. If an object forms bottom up and has a high eccentricity orbit, it appears to be an outlier; if the object forms like a ``failed star," its eccentricity is less interpretable. A low mass/low mass ratio and low eccentricity could favor a ``planetary" formation channel via either core accretion or disk instability, whereas a moderate mass and lower eccentricity could likewise favor a disk instability formation, and a higher mass and/or eccentricity could favor a core fragmentation interpretation, the ``failed-binary" interpretation.

% \par Additionally, the low age dispersion of the association and well measured parallax for the system make it an excellent addition to the small but growing landscape of ``benchmark" brown dwarfs, which are now helping to validate and constrain interior, evolutionary, and atmospheric models of brown dwarfs. 
\par Here, we present new interferometric measurements of the companion from VLTI/GRAVITY, including orbital coverage through periastron passage. GRAVITY has been used in the past few years to observe, for the first time with long-baseline optical interferometry, a handful of exoplanets and substellar companions: HR~8799~e \citep{GRAVITYCollaboration2019}, $\beta$~Pic~b and c,\citep{GravityCollaboration2020, Lagrange2020, Nowak2020, Lacour2021}, HD~206893~B and c \citep{Kammerer2021, Hinkley2023}, PDS~70~b and c \citep{Wang2021}, HD~79246~B \citep{Balmer2023}, and HIP~65426~b \citep{Blunt2023}. On high contrast targets, GRAVITY consistently achieves 50-100 \textmu as precision astrometry, while providing valuable spectral information about carbon-bearing molecules in the K-band, which can be used to constrain atmospheric modeling \citep{GravityCollaboration2020, Kammerer2021, Blunt2023}, in particular atmospheric retrievals \citep{Molliere2020, Balmer2023}. We use our new measurements of HD~136164~Ab along with the proper motion measurements of the host star from the HGCA to derive a new dynamical mass estimate. We detect absorption due to carbon monoxide in the companion's K-band spectrum. The well determined properties of this young brown dwarf companion make it an excellent prospect for future observations, and additional tests of evolutionary models.

\section{Observations} \label{sec:obs}
\subsection{VLTI/GRAVITY} \label{subsec:gravity}
\par We observed HD~136164~A and Ab four times between 2022-02 and 2023-05 using the GRAVITY instrument \citep{GravityCollaboration2017} at the European Southern Observatory (ESO) Very Large Telescope Interferometer (VLTI). We used the four 8.2m Unit Telescopes (UTs) in dual-field on-axis fringe tracking mode \citep{Lacour2019}, where the fringe tracking fiber was placed at the location of the host star HD~136165~A and used to track the fringe pattern, and the science fiber was placed at the predicted location of the companion and integrated to detect the shifted fringe pattern from the companion. Observations on 2022-02-19 and 2023-05-10 were taken as target visibility or bad weather backups to Program 1104.C-0651 (PI: Lacour) in Visitor or designated Visitor Mode (dVM). Observations on 2022-04-21 and 2022-06-14 were taken as part of Program 109.237J.001 (PI: Balmer) in Service Mode (SM). Table \ref{tab:obslog} records our observing log.
\par During the first observation of the companion with GRAVITY on 2022-02-19, its position was relatively uncertain. The fibre placement was made based on a preliminary orbit fit to the two previous SPHERE observations \citep{Wagner2020} taken 2 years before the initial GRAVITY observation, and therefore the first GRAVITY observation suffered from fiber injection losses on the companion (with a theoretical coupling efficiency of $\gamma=0.88<1$, as opposed to $\gamma>0.95$ for subsequent observations). 
\begin{figure*}
    \centering
    \includegraphics[width=\textwidth]{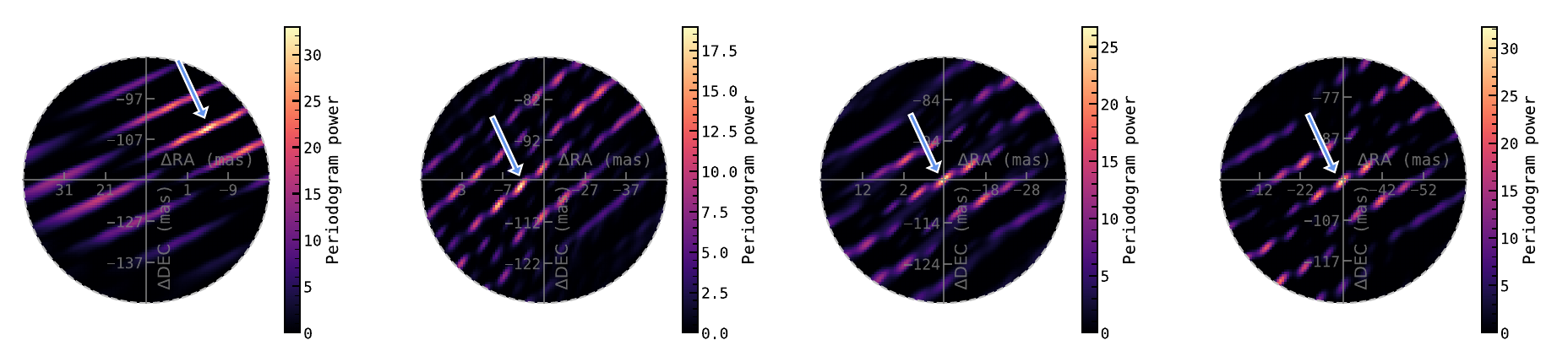}
    \caption{Detections of HD~136164~Ab with VLTI/GRAVITY. Each panel visualizes the $\chi^2$ map calculated after subtracting the stellar contamination. Each epoch in Table \ref{tab:obslog} is presented chronologically, left to right. The dashed grey circle indicates the fiber field-of-view. The origin is the placement of the science fiber on-sky for a given observation, a prediction based on the previous available orbit fit, and the units are in $\Delta$ RA and Dec, e.g. the displacement of the fiber with respect to the host star HD~136164~A. The strongest peak in the $\chi^2$ map (indicated with a blue arrow) reveals the position of the companion, with characteristic interferometric side-lobes whose shape and distribution depend on the u-v plane coverage.}
    \label{fig:detections}
\end{figure*}
\par We determined the complex visibilites on the host and the companion, which were phase-referenced with the metrology system, using the Public Release 1.5.0 (1 July 2021\footnote{\url{https://www.eso.org/sci/software/pipelines/gravity/}}) of the ESO GRAVITY pipeline \citep{Lapeyrere2014}. We proceeded to decontaminate the science fiber flux due to the halo from the host star by simultaneously fitting the stellar contamination as a low-order polynomial, and the companion as a point source. This pipeline is described in detail in Appendix A of \citet{GravityCollaboration2020}. 

\par We determined the relative astrometry of the companion by analysing the phase of the ratio of coherent fluxes. We calculate a $100\times100\,\chi^2$ periodogram power map over the fiber's field-of-view (Figure \ref{fig:detections}), and take the minimum of the $\chi^2$ map as the preliminary companion position. We then calculate another $100\times100\,\chi^2$ grid with a range restricted to $\pm10\,\mathrm{mas}$ around the initial $\chi^2$ grid minimum, and we estimate the uncertainty and co-variance of the measurement by computing the RMS of the $\chi^2$ minima for each exposure. Because HD~136164~Ab is relatively bright we only observe 3 exposures per epoch. In order to estimate our errors, we divide each exposure into four subsections, each 4 detector integrations long, for a total of 12 $\chi^2$ grid computations, except for 2023-05-10, where we only have 2 exposures with 3 detector integrations each, and 6 $\chi^2$ grid computations. We find the typical precision from this technique is on the order of $\sim\!100\,\mathrm{\mu as}$, greater than  $16.5\,\mathrm{\mu as}$ (the theoretical limit of VLTI/GRAVIY), due to systematic phase errors\footnote{In future work we intend to explore the cost/benefit of using a dynamic nested sampler, as opposed to a $\chi^2$ grid, to estimate the astrometry.} Once the astrometry is determined, we extract the ratio of the coherent flux between the two sources at the estimated position of the companion, calculating the ``contrast spectrum" of the companion.

\begin{deluxetable*}{ccccccccccc}
\tablewidth{\textwidth}
\tablecaption{Observing log. NEXP, NDIT, and DIT denote the number of exposures, the number of detector integrations per exposure, and the detector integration time, respectively, and $\tau_0$ denotes the atmospheric coherence time. The fiber pointing is the placement of the science fiber relative to the fringe tracking fiber (which is placed on the central star), $\gamma$ is the coupling efficiency at the position of the companion (see Table \ref{tab:astrometry}). \label{tab:obslog}}
\tablehead{Date & \multicolumn{2}{c}{UT time} & \multicolumn{2}{c}{NEXP/NDIT/DIT} & Airmass & $\tau_0$ & Seeing & Fiber pointing & $\gamma$ \\
& Start & End & HD~136264~Ab & HD~136264~A & & & & $\Delta$RA/$\Delta$DEC &}
\startdata
\hline\hline
2022-02-20 & 08:50:14 & 09:25:14 & 4/12/30~s & 4/64/1~s & 1.03-1.06 & 9.5-12.3~ms & $0.37-0.99^{\prime\prime}$ & 11/-117  & 0.878  \\
2022-04-21 & 05:16:58 & 05:55:48 & 4/12/30~s & 4/64/1~s & 1.03-1.06 & 1.8-2.9~ms & $0.95-1.53^{\prime\prime}$ & -8/-103  & 0.999 \\
2022-06-15 & 03:36:41 & 04:16:26 & 4/12/30~s & 4/64/1~s & 1.05-1.12 & 1.8-3.1~ms & $0.95-1.37^{\prime\prime}$ & -17/-101  & 0.979 \\
2023-05-10 & 04:24:53 & 04:40:20 & 2/6/30~s & 2/48/1~s & 1.02-1.03 & 7.9-11.1~ms & $0.40-0.53^{\prime\prime}$ & -32/-97  & 0.999 \\
\hline
\enddata
\end{deluxetable*}

\vspace{-0.35in}
\par Observations from 2022-02-19, 2022-04-21, and 2022-05-10 yield robust detections of the companion. Due to a metrology glitch during the 5th DIT in the observing sequence on 2022-06-14, the baselines for UT1 are unrecoverable for about half of the observing sequence, and we neglect these baselines in our extraction. We nevertheless recover the companion's signal in this data, albeit with larger astrometric uncertainties (of the same magnitude as the original SPHERE observations). Figure \ref{fig:detections} illustrates the detections of HD~136164~Ab within the fiber field-of-view. 
\par
\begin{figure*}
    \centering 
    \includegraphics[width=0.85\textwidth]{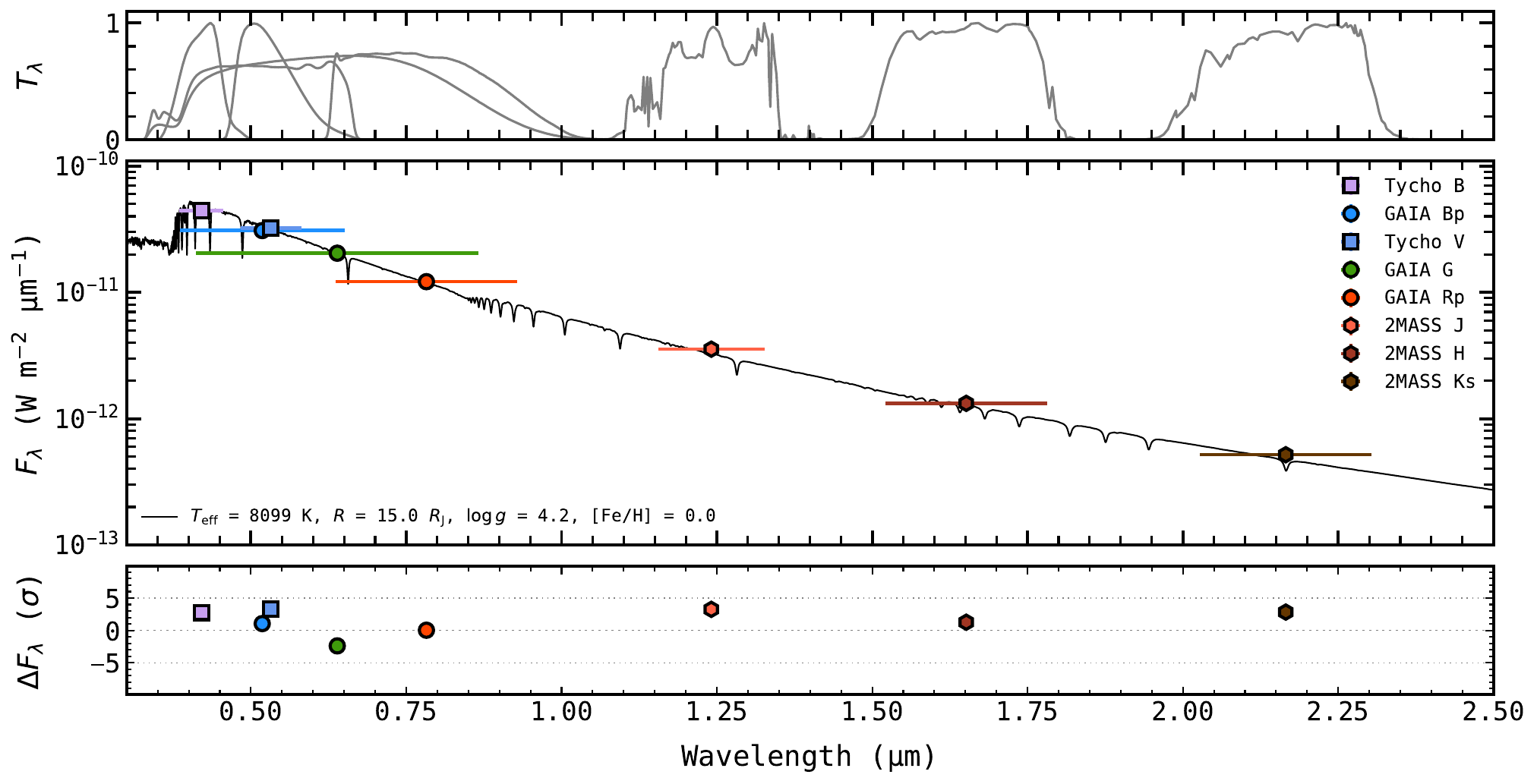}
    \caption{The SED of HD~136164~A. Archival photometry is overplotted a \texttt{BT-NextGen} model spectrum with $\mathrm{T_{eff}}=8100\,\mathrm{K}$ and $\log(g)=4.2$. This spectrum is used to transform the companion's contrast spectra into flux calibrated spectra.}
    \label{fig:host}
\end{figure*}

 We transformed our contrast spectrum of the companion into a flux calibrated spectrum using a synthetic spectrum of the host star. We scaled a \texttt{BT-NextGen} \citep{Allard2011} spectrum with $\mathrm{T_{eff}}=8100\,\mathrm{K}$ and $\log(g)=4.2$ \citep{Bochanski2018} to archival photometry from \textit{Gaia} \citep{GaiaCollaboration2022}, Tycho2 \citep{Hog2000}, and 2MASS \citep{Cutri2003} using \texttt{species} \citep{Stolker2020}. We broadened the spectrum with \texttt{specutils} \citep{astropy:2013, astropy:2018, astropy:2022} using a Gaussian kernel with a FWHM of 2~\AA~to match the rotational broadening of the star, and then sampled the broadened spectrum on the SPHERE and GRAVITY wavelength grids using \texttt{spectres} \citep{Carnall2017}. The spectrum and stellar photometry is shown in Figure \ref{fig:host}.

\begin{deluxetable*}{RRRRRR}
\tablewidth{\textwidth}
\tablecaption{New relative astrometry of HD~136264~Ab around HD~136264~A. \label{tab:astrometry}}
\tablehead{
\multicolumn{6}{c}{GRAVITY} \\ 
\colhead{Epoch [MJD]} & \colhead{$\Delta$RA [mas]} & \colhead{$\sigma_{\Delta\mathrm{RA}}$ [mas]} & \colhead{$\Delta$Dec [mas]} & \colhead{$\sigma_{\Delta\mathrm{Dec}}$ [mas]} & \colhead{$\rho$}
}
\startdata
59630.38 & -3.92 & 0.07 & -104.19 & 0.05 & -0.67 \\
59690.24 & -7.87 & 0.04 & -103.45 & 0.07 & -0.68 \\
59745.17 & -11.55 & 0.43& -103.38 & 0.79 & 0.53 \\
60074.19 & -32.12 & 0.04 & -97.64 & 0.05 & -0.69 \\
\hline\hline
\enddata
\tablecomments{The co-variance matrix can be reconstructed using $\sigma_{\Delta\mathrm{RA}}^2$ and $\sigma_{\Delta\mathrm{Dec}}^2$ on the diagonal, and $\rho\times\sigma_{\Delta\mathrm{RA}}\times\sigma_{\Delta\mathrm{Dec}}$ on the off-diagonal.}
\end{deluxetable*}

% \subsection{JWST/NIRCam}

\subsection{VLT/SPHERE} \label{subsec:sphere}

\par HD~136164~Ab was first detected using the Spectro-Polarimetic High contrast imager for Exoplanets REsearch \citep[SPHERE][]{Beuzit2019} at the ESO Very Large Telescope (VLT). We adopt the companion's 2015-06-19 and 2019-06-29 astrometry from \citet{Wagner2020} as-is, although we note that, when compared with our updated orbital fits including our GRAVITY observations, the observation on 2019-06-29 appears systematically biased inwards in separation, likely because the off-axis PSF becomes asymmetric due to the transmission function of the coronagraph at close separations. Because our GRAVITY observations drive the orbital fits, and the 2015-06-19 astrometry is unaffected by this systematic error, we do not correct the SPHERE astrometry in this work. 
\par We re-reduced the IFS data from 2015-06-19 in order to extract the spectrum of the companion. The data were taken in the \texttt{IRDIFS-EXT} mode where the IFS \citep{Claudi2008} and IRDIS \citep{Dohlen2008} observe in parallel. The IFS covers wavelength bands Y, J, and H1 between $0.95-1.65\,\mu\mathrm{m}$ and the dual band imager IRDIS covers K1 and K2 at $2.0\,\mu\mathrm{m}$ and $2.1\,\mu\mathrm{m}$ \citep{Vigan2010}. Because our GRAVITY data cover the K-band, we do not re-reduce the IRDIS data. The observations used the \texttt{N\_ALC\_YJH\_S} coronagraph, which is optimized for close inner working angles. The raw data from the ESO archive were pre-processed with \texttt{vlt-sphere}, an open-source pipeline \citep{Vigan2020}, and then we performed starlight subtraction using \texttt{pyKLIP} \citep{wang2015}, an open source implementation of Karhunen-Lo\`eve Image Processing \citep{Soummer2012}. We used the forward modeling capabilities of KLIP \citep{Pueyo2016} to extract the spectrum of the companion using \texttt{pyKLIP} \citep{Wang2016, Greenbaum2018}, and applied the theoretical coronagraph transmission function to the resulting spectrum. The contrast spectrum was flux calibrated as in \S\ref{subsec:gravity}.

% \subsection{SOAR/TripleSpec4.1}

\section{Analysis} \label{sec:analysis}

\subsection{Orbital analysis} \label{subsec:orbit}

\par In order to measure the orbit of HD~136164~A and b, we use \texttt{orbitize!}\footnote{\href{https://orbitize.readthedocs.io/en/latest/}{orbitize.readthedocs.io}.} \citep{Blunt2020}. We use \texttt{orbitize!}'s parallel-tempered \citep{Vousden2016} Affine-invariant \citep{Foreman-Mackey2013} MCMC algorithm. This algorithm fits for the 6 parameter visual orbit \citep{Green1985}, the system parallax, the proper motions and reference epoch astrometry, and the masses of the star and companion. We used 1,000 walkers at 20 temperatures to sample our posterior distribution of orbits, with 15,000 steps of burn-in discarded, before 5,000 steps were recorded (for a total of 50,000 orbits recorded in our posterior distribution). We place a physically motivated normally distributed prior $\mathcal{N}(1.7,\,0.1)\,\mathrm{M_\odot}$ \citep{Bochanski2018, Kervella2022}, and a log-uniform prior on the companion mass of $\log\,\mathcal{U}(1,200)\,\mathrm{M_J}$, a Gaussian prior on the system parallax based on the \textit{Gaia} DR3 measurement $\mathcal{N}(8.202, 0.040)\,\mathrm{mas}$ and otherwise use default priors on all orbital elements as described in \citet{Blunt2020}. We fit the model to relative astrometry from SPHERE \citep{Wagner2020}, GRAVITY (this work), and absolute astrometry on the host star from the \textit{Hipparcos}-\textit{Gaia} Catalogue of Accelerations \citep{Brandt2021}. 

\par
\begin{figure*}
    \centering
    \includegraphics[width=\textwidth]{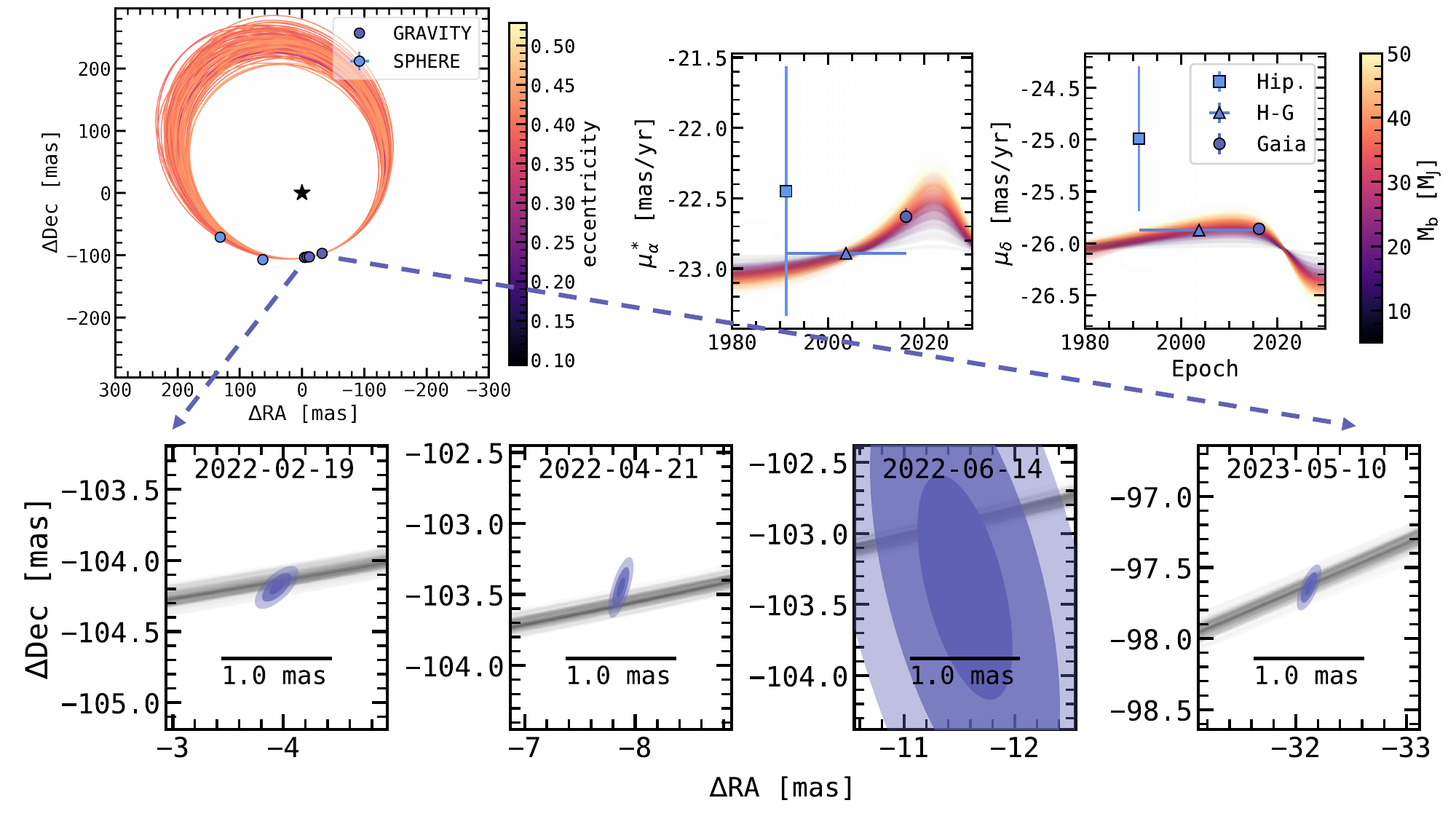}
    \caption{HD~136164 orbital analysis. \textbf{Top left:} The visual orbit of HD~136164~Ab around HD~136164~A (black star). 1,000 randomly chosen orbits from our fit using the MCMC are shown in color, with the color corresponding to the eccentricity of a given orbit. \textbf{Top middle and right:} The proper motion of the host star HD~136164~A over time in RA and Dec (middle and right, respectively). The proper motion of the host as predicted by the orbit fit is shown in curves colored corresponding to the mass of the companion. Observations from \textit{Hipparcos} and \textit{Gaia} are shown with errorbars as recorded in the HGCA. Note that the \texttt{orbitize!} proper motion log-likelihood function evaluates the time-averaged proper motions, and not the instantaneous proper motions that are over-plotted here \citep[see][]{Hinkley2023}; the instantaneous measurements are shown here for visualization purposes only. \textbf{Bottom:} VLTI/GRAVITY measurements of the position of HD~136164~Ab across four epochs (see Table \ref{tab:astrometry}), plotted against the same sample of MCMC orbits as in the top left panel. The $1$, $2$, and $3\,\sigma$ confidence contours on each measurement are shown with decreasing transparency; the covariance of the measurement is related to the u-v plane coverage of the observation.}
    \label{fig:orbit}
\end{figure*}

 Figure \ref{fig:orbit} plots the best fit orbits to the observations of the system, and Figure \ref{fig:orbit-post-full} illustrates the posterior distribution of best fit orbits from our \texttt{orbitize!} MCMC. We note that the 2nd SPHERE epoch deviates by a few milliarcseconds from our orbit fits, likely because the coronagraphic transmission at the location of the companion distorts the photocenter, biasing the astrometry in separation. We find a highly eccentric orbit, with $e=0.44\pm0.03$, and derive a companion mass of $35\pm10\,\mathrm{M_J}$. Our derived stellar mass ($1.87\pm0.07\,\mathrm{M_\odot}$) deviates by $1\,\sigma$ from our prior on the stellar mass, resulting in a mass ratio $q=0.02$. Our four epochs of relative astrometry at $<100\,\mu\mathrm{as}$ precision, combined with the long time baseline information from the absolute astrometry, has strongly constrained the visual orbit of the companion.

\begin{deluxetable*}{ccccc}
\tablewidth{\textwidth}
\tablecaption{Orbital parameters inferred for HD~136164~Ab in this work. \label{tab:orbit}}
\tablehead{
\colhead{Parameter} & \colhead{Prior} & \colhead{Median} & \colhead{Lower 1$\,\sigma$ CI} & \colhead{Upper 1$\,\sigma$ CI}
}
\startdata
a [au]                         & $\log\,\mathcal{U}(0.001, 1e4)$                               & 22.48 & -1.03 & 1.15 \\
e                              & $\mathcal{U}(0,1)$                                   & 0.44 & -0.03	& 0.03 \\
i [rad]                        & $\mathrm{Sine}(0,\pi)$                                    & 0.20	& -0.09 & 0.08 \\
$\omega$ [rad]                 & $\mathcal{U}(0,2\pi)$                         & 2.24 & -0.53 & 0.43 \\
$\Omega$ [rad]                 & $\mathcal{U}(0,2\pi)$                    & 1.31 & -0.43 & 0.47 \\
$\tau$ [dec. yr]$^\dag$          & $\mathcal{U}(0,\mathrm{P})$ & 2023.67 & -0.21 & 0.36\\
$\pi$ [mas]                         & $\mathcal{N}(8.202, 0.040)$               & 8.23 & -0.04	& 0.04 \\
$\mathrm{M_B}$ [$\mathrm{M_J}$] & $\log\,\mathcal{U}(1,200)$                                     & 34.9 & -9.7	& 9.9 \\
$\mathrm{M_A}$ [$\mathrm{M_\odot}$] & $\mathcal{N}(1.7,\,0.1)$                                      & 1.8725 & -0.06 & 0.07
\enddata
\tablecomments{We report the median and 68\% confidence interval on each parameter derived from the posterior visualized in Figure \ref{fig:orbit-post-full}. Propogating errors between both mass estimates yields $q=0.018\pm0.005$. $^\dag$Next periastron passage after $\tau_{\mathrm{ref}}=2020.0$}
\end{deluxetable*}

\subsection{Spectral analysis} \label{subsec:spectrum}
\begin{figure*}
    \centering
    \includegraphics[width=0.9\textwidth]{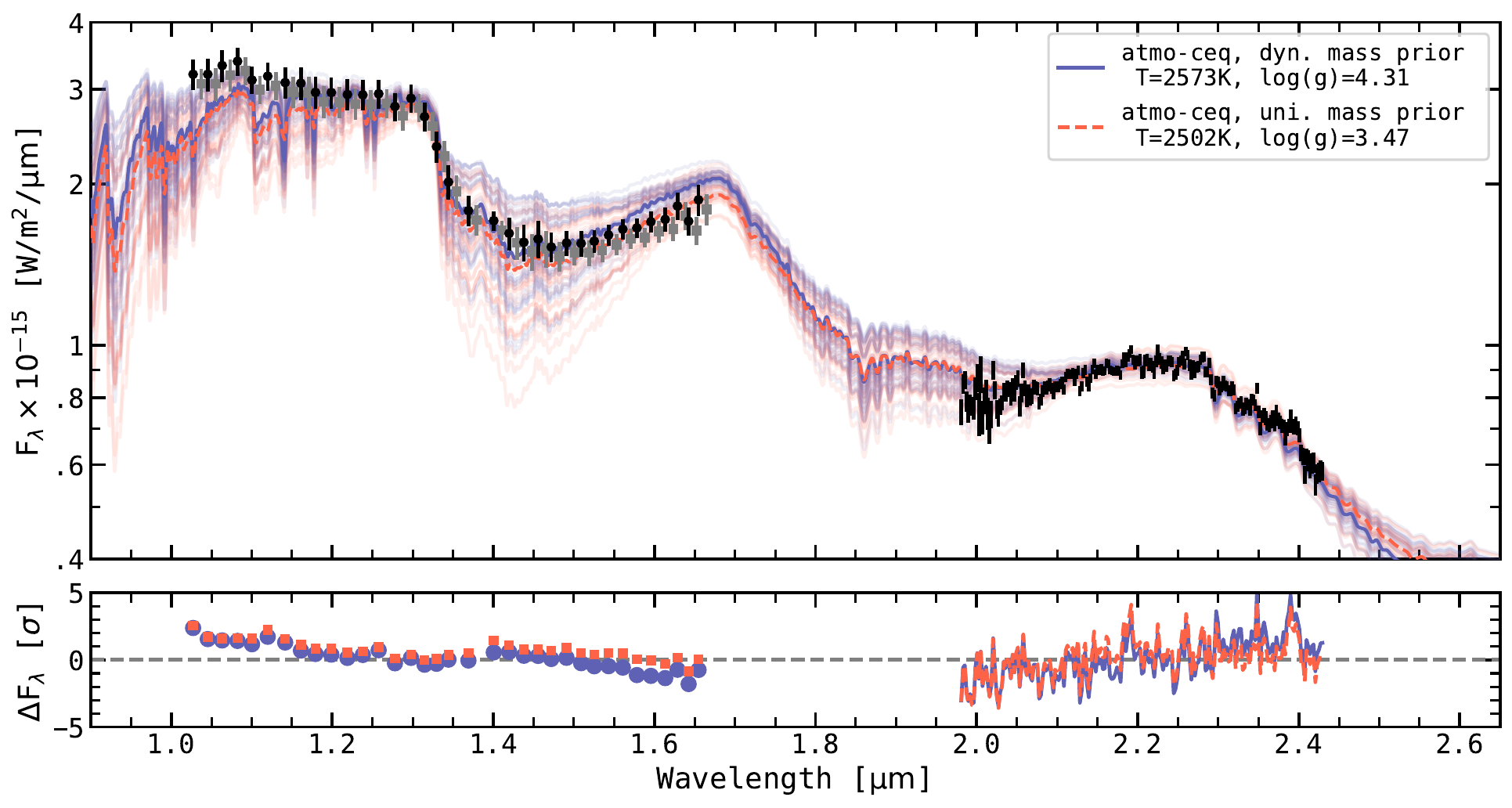}
    \caption{The spectrum of HD~136164~Ab (black data points) fit to the cloudless \texttt{ATMO} equilibrium chemistry self-consistent model, with and without a prior on the dynamical mass of the companion (median samples are plotted as blue and red curves, respectively). 30 random samples from each posterior distribution are plotted as semi-transparent curves. In each fit, the SPHERE YJH1 data is allowed to scale by a factor, so for the fit with a dynamical mass prior, the scaled SPHERE data is shown as black circles, and without a dynamical mass prior, shifted to the right for visualization as gray squares.}
    \label{fig:atmo-fit}
\end{figure*}

\par We compared the spectrum of the companion to two self-consistent, cloudless, chemical equilibrium model grids, \texttt{ATMO} \citep{Phillips2020} and \texttt{SPHINX} \citep{Iyer2023}. We chose these model grids because they use modern opacity sources and are computed for temperature ranges ($2300-3000~\mathrm{K}$) appropriate for the spectral type of HD~136164~Ab. In particular, the \texttt{SPHINX} grid accounts for spectral broadening of key molecules, a range of metallicities and C/O ratios, and has been benchmarked to observations of M-dwarfs with known bulk properties. However, it is computed for a lower resolution ($R=250$) than our GRAVITY data ($R=500$), so when fitting the grid to our data, we convolve the GRAVITY spectrum to the resolution of the grid using \texttt{spectres} \citep{Carnall2017}. 
\par We used the \texttt{species}\footnote{\href{https://species.readthedocs.io}{species.readthedocs.io}} package to fit our spectra to the model grid, linearly interpolating spectra between grid points. In version 0.71 of \texttt{species}, we ingested the \texttt{SPHINX} from \citep{iyer2022_data} that assumes a mixing length parameter of 1. We initialized \texttt{pyMultiNest}\footnote{\href{https://johannesbuchner.github.io/PyMultiNest/}{johannesbuchner.github.io/PyMultiNest/}} \citep{Feroz2008, Feroz2009, Buchner2014} via \texttt{species} to sample the interpolated grid with 500 live points. We measured the posterior distribution on the grid parameters, namely effective temperature ($\mathrm{T_{eff}}$), log(g), radius, parallax, [Fe/H] and C/O for the \texttt{SPHINX} grid, and included a Gaussian process parameterized by a squared-exponential kernel to account for correlated noise between wavelength channels in the SPHERE data \citep[see \S4.1 in][]{Wang2020}. When fitting the GRAVITY data, \texttt{species} accounts for the correlation matrix of the spectrum in the fit. We weight each spectral point across the entire SED according to its wavelength spacing so that the GRAVITY spectrum, with an order of magnitude more data points, does not completely dominate the fit. To account for an apparent offset in the flux of the two spectra, we fit for a scaling parameter on the SPHERE YJH1 spectrum, as is common in the literature \citep[e.g.][]{Molliere2020}. We place the same parallax prior on the atmospheric fits as we do the orbital fits in \S\ref{subsec:orbit}. 
\par We fit each model grid to our data twice, first with a uniform prior on the mass (related to the sampled parameters $\log(g)$ and radius), and second with a Gaussian prior on the mass corresponding to the dynamical mass from our orbit fit in \S\ref{subsec:orbit}, $\mathcal{N}(35,10)~\mathrm{M_J}$. This experiment allows the sampler to determine the log(g) and radius based solely on the spectral and parallax measurements, and then adds the additional constraint of the dynamical mass. Figure \ref{fig:atmo-fit} plots the spectrum of the companion and the results of the \texttt{ATMO} model fitting. We find the median sample has  $\mathrm{T_{eff}}$, $\log(g)$, $\mathrm{R_p}$, $\mathrm{\log{(L/L_\odot)}}$ and SPHERE scale factor of $2530\,\mathrm{K}$, $3.5$, $2.0\,\mathrm{R_J}$, $-2.8$, $0.70$, respectively, with a uniform mass prior, and $2600\,\mathrm{K}$, $4.3$, $1.9\,\mathrm{R_J}$, $-2.8$, $0.71$ with the dynamical mass prior. The residuals to the fit are strongly wavelength dependent, and appear related to regions where water and iron-hydride opacity is dominant in shaping the spectral slope, especially when the dynamical mass prior is considered. This could indicate model deficiencies (non-solar abundances, reduced temperature gradients, or dis-equilibrium chemistry) or data deficiencies (improperly corrected tellurics, phase errors, or coronagraphic effects).

\begin{figure*}
    \centering
    \includegraphics[width=0.9\textwidth]{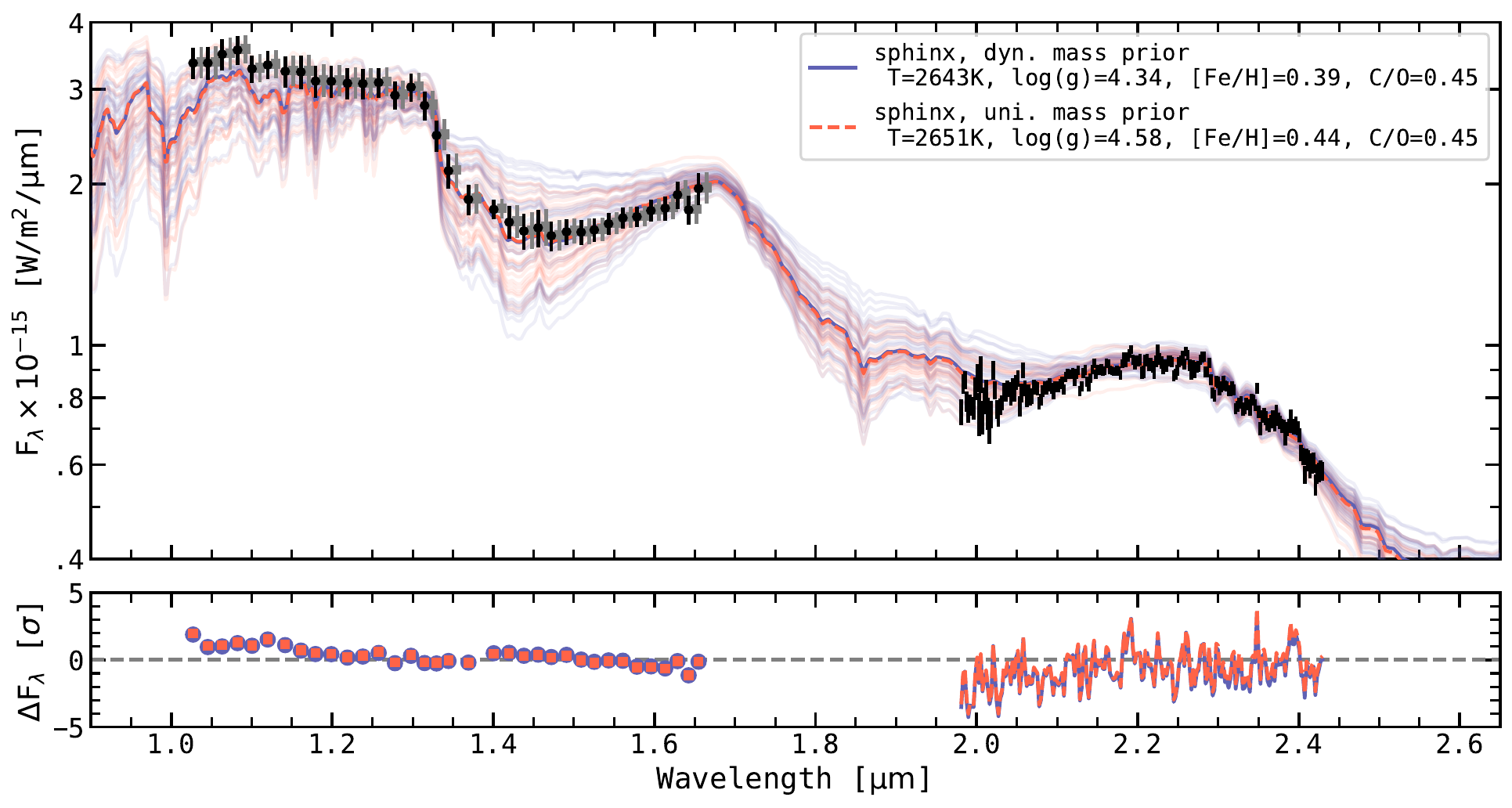}
    \caption{The spectrum of HD~136164~Ab (black data points) fit to the cloudless \texttt{SPHINX} equilibrium chemistry self-consistent model, which varies abundance in addition to effective temperature and surface gravity, with and without a prior on the dynamical mass of the companion (median samples are plotted as blue and red curves, respectively). 30 random samples from each posterior distribution are plotted as semi-transparent curves. In each fit, the SPHERE YJH1 data is allowed to scale by a factor, so for the fit with a dynamical mass prior, the scaled SPHERE data is shown as black circles, and without a dynamical mass prior, shifted to the right for visualization as gray squares.}
    \label{fig:sphinx-fit}
\end{figure*}

Figure \ref{fig:sphinx-fit} plots the spectrum of the companion and the results of the \texttt{SPHINX} model fitting. We find the median sample has  $\mathrm{T_{eff}}$, $\log(g)$, $\mathrm{R_p}$, [Fe/H], C/O, $\mathrm{\log{(L/L_\odot)}}$, and SPHERE scale factor of $2650\,\mathrm{K}$, $4.6$, $1.8\,\mathrm{R_J}$, $0.42$, $0.47$, $2.8$, and $0.74$, respectively, with a uniform mass prior, and $2640\,\mathrm{K}$, $4.35$, $1.9\,\mathrm{R_J}$, $0.39$, $0.45$, $2.8$, and $0.74$ with the dynamical mass prior. The residuals to the fit are smaller compared to the fit to the model assuming solar abundances (Figure \ref{fig:atmo-fit}), and the impact of the dynamical mass prior is more subtle. With more free parameters, the model can produce an equivalently good fit to the data with or without the dynamical mass prior, and no longer appears biased towards an implausibly low surface gravity without the dynamical mass prior.

\subsection{Evolutionary model comparison}

\par As a byproduct of our spectral fits, we use \texttt{species} to integrate under the sampled \texttt{ATMO} and \texttt{SPHINX} spectra and determine the distributions of bolometric luminosity for both the uniform and dynamical mass prior cases. The \texttt{ATMO} model gives $\mathrm{\log{(L/L_\odot)}}=-2.8\pm0.3$ for both cases, and the \texttt{SPHINX} model gives $\mathrm{\log{(L/L_\odot)}}=-2.8\pm0.3$ for both cases.

\par For substellar objects, there is a frequently observed tension between the evolutionary and spectral effective temperatures \citep[see, e.g.][their Figure 23]{Sanghi2023}. We leverage our dynamical mass to compare the bolometric luminosity expected for a $16\pm2\,\mathrm{Myr}$ brown dwarf according to the \texttt{ATMO2020} evolutionary model \citep{Phillips2020} and the bolometric luminosity we derive from our spectral fits. The model uses the \texttt{ATMO} atmospheric model as a surface boundary condition to calculate the interior structure and evoltion of substellar objects.
\par We drew luminosities for a $35\,\mathrm{M_J}$ brown dwarf at $14$, $16$, and $18\,\mathrm{Myr}$, finding $\mathrm{\log{(L/L_\odot)}_{evol}}=-2.68^{+0.08}_{-0.05}$. Given the larger uncertainties on our spectrally derived bolometric luminosity (not to mention the potential systematic uncertainty on this estimate, given that we must rely on atmospheric models to extrapolate beyond the $1-2.5\,\mathrm{\mu m}$ range), the evolutionary and spectral bolometric luminosities for HD~136164~Ab agree to within $1\,\sigma$. Increasing the wavelength coverage on the companion (especially at shorter wavelengths) will decrease the uncertainty on the spectrum-derived luminosity, and could reveal a tension if it exists. These models reveal the un-physical nature of the spectrally derived $\log(g)$ in the absence of a dynamical mass prior.

\par
\begin{figure}
    \centering
    \includegraphics[width=0.45\textwidth]{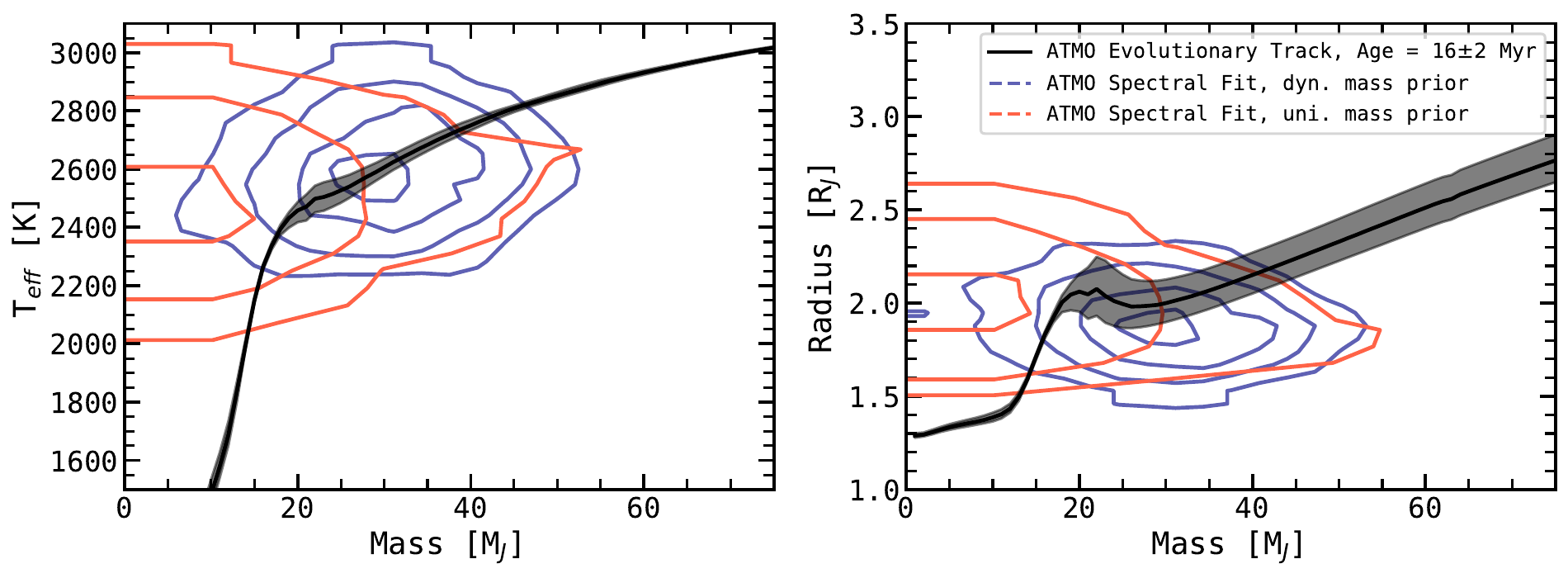}
    \caption{\textbf{Left:} Mass versus effective temperature for our atmospheric fits (contours) compared to the mass radius relationship from the \texttt{ATMO2020} evolutionary model (black curve). \textbf{The uncertainty on the system's age is represented by gray contours to the black curve, $16\pm2\,\mathrm{Myr}$.} When the dynamical mass is included as a prior in the atmospheric model fit, the mass and effective temperature are consistent with the evolutionary model prediction to within $1\,\sigma$. \textbf{Right:} Mass versus radius for our atmospheric fits compared to the mass radius relationship from an evolutionary model. Without a dynamical mass prior, our spectral fits derive a $\log(g)$ that results in a mass inconsistent with expectations from the evolutionary model. With a dynamical mass prior, the determined $\log(g)$ is consistent with expectations to within $1\,\sigma$.}
    \label{fig:mass_radius}
\end{figure}

 Figure \ref{fig:mass_radius}a plots the distributions on mass and radius from our atmospheric fits, compared to the \texttt{ATMO2020} mass-radius model \citep{Phillips2020} for an age of $16\pm2$, showing that these parameters are in agreement within $1-2\,\sigma$ with expectations from the evolutionary model. Figure \ref{fig:mass_radius}b plots the distributions on mass and effective temperature. When the dynamical mass is included as a prior in the spectral fit, the temperatures are consistent within $1\,\sigma$. For many brown dwarfs, there is significant tension between the evolutionary and spectral effective temperatures. 

\section{Discussion} \label{sec:discuss}
\subsection{The formation of HD~136164~Ab}
\par We have demonstrated that HD~136164~Ab has an eccentric orbit ($e=0.44\pm0.03$), a dynamical mass of $35\pm10\,\mathrm{M_J}$, and a mass ratio with HD~136164~A of $q=0.018\pm0.005$ (see \S\ref{subsec:orbit}). This makes HD~136164~A the youngest substellar companion with a dynamical mass estimate. The dynamical mass agrees within uncertainties with the previous evolutionary model derived mass based on SPHERE spectrophotometry, and are in perfect agreement when considering the evolutionary models anchored to the SPHERE/IRDIS K-band photometry \citep{Wagner2020}. The dynamical mass and mass ratio is inconsistent with core accretion models of formation \citep[e.g.][]{Mordasini2009, Emsenhuber2021a, Emsenhuber2021b}. The good agreement with the dynamical mass, spectral effective temperature, and the \texttt{ATMO} isochrone with a Sco-Cen age (see Figure \ref{fig:mass_radius}) can be interpreted as evidence for the similarity in age of the host and companion, the coevality of the two, and therefore additional evidence against core accretion.
\par Simulations of disk-fragment formation and fragment-fragment interactions in \citet{Forgan2013, Forgan2015, Forgan2018} appear to indicate the result of disk instability formation is a distribution of single giant planets/brown dwarfs near the deuterium burning limit at wide separations ($10-100\,\mathrm{au}$), with a distribution peaking at low orbital eccentricities, but flattening towards a non-zero relative frequency between $e=0.1-0.8$. This could tentatively support the interpretation that HD~136164~Ab formed via disk instability, since it is a widely separated companion at $22.5\,\mathrm{au}$. It could be that the companion formed from disk fragmentation with an initially low eccentricity that was then increased via dynamical interactions with the outer M-dwarf HD~136164~B at $650\,\mathrm{au}$, but this would take time to evolve dynamically, and the system is young. Demonstrating this would require dynamical modeling and additional orbital monitoring of B and Ab. We could interpret the slightly enhanced metallicity we derive for the companion using the \texttt{SPHINX} models as evidence the companion accreted a significant fraction of metals from the circumstellar environment, if it indeed formed via disk instability. For instance, it could have undergone late accretion of solids \citep[e.g.][]{Nowak2020}. The posterior distribution on abundances from our model fits to the \texttt{SPHINX} grid are wide, and still encompass solar metallicity solutions at the $2-3\,\sigma$ level. In \citet{Wagner2020}, it was suggested that a low C/O ratio could imply an earlier period of gas accretion from the disk, because during the Class 0 stage of the star/circumstellar evolution, no CO gas would be frozen out, whereas during the Class 1 stage, CO gas would be frozen out beyond $30\,\mathrm{au}$. Our best fit \texttt{SPHINX} spectrum has a lower C/O ratio of $0.45$, although the posterior still spans a wide range of C/O encompassing the solar value at $2-3\,\sigma$, and similarly this could be loosely interpreted as evidence for formation of the companion within the circumstellar disk. The host star could also have a sub-solar C/O abundance, and the lower C/O abundance for HD~136164~Ab could be reflective of a match to the stellar abundance.
\par \citet{Kratter2010} show that disk fragmentation can produce brown dwarf and even planetary mass companions, but provides the additional interpretation that these systems are necessarily ``failed stars," i.e. that the distribution of disk instability born companions has a low mass tail, but increases above the deuterium burning limit. They predict an increase in the occurrence of brown dwarf companions to A-type stars; as is the case here, but also appears to be the case more generally \citep{Nielsen2019}.
\par If it did not form from a fragmenting circumstellar disk, could HD~136164~Ab have formed via the collapse of a molecular cloud, as a product of the IMF? With opacity limited minimum mass estimates as low as $1-3\,\mathrm{M_J}$ for core fragmentation \citep[e.g.][]{Boyd2005, Bate2009}, or a derived minimum mass of around $6-18\,\mathrm{M_J}$ from radiation hydrodynamical simulations \citep[e.g.][]{Bate2012}, the mass of HD~136164~Ab is not unprecedented for a by-product of the fragmenting core mass function. In \citet{Bate2012}, the outcome of the stellar IMF produces a handful of binaries with small mass ratios ($0.01-0.02$) with a relatively uniform distribution of eccentricities. HD~136164~Ab's high eccentricity and mass ratio of $0.02$ could therefore indicate a core fragmentation formation pathway, if eccentricities are expected to be damped during formation in a disk, but not during core fragmentation. In the context of population level studies of substellar companions, this eccentricity also empirically indicates more of a ``brown dwarf/failed star" nature, as directly imaged planets appear to follow a distinct eccentricity distribution with generally lower eccentricities (as their eccentricities are expected to be damped by their parent disks), whereas substellar objects produced by fragmenting cores tend to have a distribution of on average higher eccentricities \citep{Bowler2020, Nagpal2023, DoO2023}. \citet{Parker2022} use N-body simulations of star forming regions to show that early type stars can capture single brown dwarfs or even steal planets from other lower mass systems within the first $10\,\mathrm{Myr}$ of an association's lifetime, so it is a reasonable assumption that, even if the brown dwarf did not form directly into a binary configuration with HD~136164~A, it could have formed nearby and been subsequently captured onto a binary orbit.
\par It is still unclear whether this brown dwarf companion could have formed from disk fragmentation or core fragmentation, but it appears clear that the companion is a ``failed star'' coeval with the host, either born or captured into a binary orbit.

\subsection{Atmospheric modeling in the context of a dynamical mass prior}
\par In \citet{Balmer2023}, we saw that the atmospheric modeling of a close separation brown dwarf companion can be strongly influenced by the inclusion of a dynamical mass prior. In that work, for a benchmark brown dwarf with a known mass and stellar abundances, the $\log(g)$ and radius of the atmospheric model appeared strongly influenced by regions of increased systematics in the spectrum. It is necessarily the case that spectral and astrometric measurements made at such close separations will suffer, at least initially, from such systematic errors. These could be responsible for both the wavelength dependent residuals between our data and the \texttt{ATMO} models. It is also the case that model deficiencies in solar abundance, chemical equilibrium, cloudless, radiative-convective equilibrium models (e.g. non-solar abundances, clouds or reduced temperature gradients, vertical mixing and disequilibrium chemistry, etc) could impact the determined physical parameters and quality of fit. The \texttt{SPHINX} grid, which varies abundances, can provide a better fit to the data, invoking a sub-solar C/O ratio, and an enhanced metallicity. In Figure \ref{fig:atmo-fit}, the residuals to the \texttt{ATMO} model occur near the edges of the instruments' respective wavelength range, and appear to strongly influence the fit $\log(g)$. In our fit without the dynamical mass prior, the posterior distribution was driven to unphysically low values, $\log(g)=3.0$. When the dynamical mass prior is considered, we find a much more reasonable $\log(g)=4.4$, and a corresponding radius of $1.9\,\mathrm{R_J}$. It could be that a circumsecondary disk or dust extinction produces the slope to the GRAVITY data. We briefly considered including an extra blackbody term, or an extinction term, in our spectral fitting to account for these effects, and were unable to convincingly constrain its presence with our data, so we do not consider that here (the disk excess could be investigated with 3-5~\textmu m JWST observations). 
\par Given the good agreement between our atmospherically derived luminosity (based on our fits to the spectrum using a dynamical mass prior) and our evolutionarily derived luminosity (based on the system age and dynamical mass), for relatively cloudless late M-type substellar companions without a dynamical mass prior, it appears reasonable to adopt an evolutionary model derived mass prior based on the observed flux of the companion when fitting the spectrum with model grids. Doing so will ensure the fit is not driven to unphysical values of $\log(g)$. We note, however, that if a fit is driven to unphysical values, it could be due to a model deficiency since, in the case of HD~136164~Ab, we found a model grid that varied abundances away from solar values derived physical $\log(g)$ values without the dynamical mass prior. As in \citet{Balmer2023}, the dynamical mass prior appears most useful in determining whether there are data or model deficiencies, and what changes (to data treatment or model construction) might alleviate the tension between the two.

\section{Conclusions}
\par In this work we:

\begin{enumerate}
    \item Observed the brown dwarf companion HD~136164~Ab four times with VLTI/GRAVITY dual-field interferometry. We detected the companion in all four observations despite limited field rotation and exposure time, demonstrating the power of GRAVITY to efficiently characterize brown dwarf companions.
    \item Extracted precise astrometry of the companion ($\mathrm{med}~\sigma=40~\mathrm{\mu as}$) and a contrast spectrum at R=500 resolution from the interferometric observables. We also re-extracted the SPHERE IFS YJH1 spectrum of the companion and flux calibrated these contrast spectra using a stellar template.
    \item Fit the orbit of the companion with a combination of relative astrometry from GRAVITY and SPHERE, and absolute astrometry from the HGCA, determining the eccentricity and dynamical mass of the companion for the first time. We leverage the moderate eccentricity ($e=0.44\pm0.03$) and mass ratio ($q=0.018\pm0.005$) to assess the formation of the object.
    \item Fit the spectrum of the companion to two grids of self-consistent, cloudless atmospheric models, \texttt{ATMO} and \texttt{SPHINX}. We discussed the inclusion of the dynamical mass as a prior in atmospheric fits for this object. The \texttt{SPHINX} grid, which varies abundances, indicates a slightly sub-solar C/O ratio ($0.45$) and enriched metallicity ($0.4$), with uncertainties that are consistent with solar values. Deriving the luminosity from our spectral fits and comparing them to expectations from evolutionary models revealed an apparent agreement, albeit with uncertainties of $0.3$ dex in luminosity.
    \item Discussed the formation of HD~136164~Ab in the context of the new GRAVITY data and especially the dynamical mass. We rule out formation via core accretion, but present evidence that could be interpreted in favor of either disk fragmentation or cloud fragmentation. 
\end{enumerate}

\par HD~136164~Ab now joins a very select group of young substellar companions with dynamical mass measurements \citep[see Figure 4 in][]{Franson2023}. It is currently the youngest substellar companion with a dynamical mass estimate. Like PZ Tel B \citep{Biller2010, Mugrauer2010, Schmidt2014, Maire2016, Stolker2020, Franson2023}, this object will be crucial for benchmarking evolutionary models in the coming decade. HD~136164~Ab does not need to stand alone, however. We have shown that, even without stellar radial velocities, coupling GRAVITY observations with the HGCA can produce well constrained dynamical mass estimates for young companions orbiting stars that are not amenable to radial velocity characterization. Future work should look to estimate the dynamical mass of the young substellar companions to HD~115470 \citep[HIP~64892~B][]{Cheetham2018}, $\mu^2\,\mathrm{Sco}$ \citep[$\mu^2\,\mathrm{Sco}$~b][]{Squicciarini2022}, HD~149274 \citep[HIP~81208~B][]{Viswanath2023, Chomez2023}, and the planet orbiting HD~116434 \citep[HIP~65426~b][]{Chauvin2017, Cheetham2018, Stolker2020, Petrus2021, Blunt2023}. While these might not result in dynamical masses of same precision as this result (since the aforementioned companions are on longer period orbits), a similar analysis might at the very least provide useful mass upper limits for assessing the formation histories of these objects.
\par For HD~136164~Ab itself, there remains a number of additional investigations that can continue to refine estimates of its bulk and atmospheric properties, to aid in the benchmarking of various models. Future observations of the companion at shorter wavelengths \citep[e.g. z' or i' band with MagAO-X,][]{Males2022}, and longer wavelengths (e.g. with JWST/NIRCam at 3-5\,\textmu m, via GO program 1902, PI: Kammerer) will produce a precise bolometric luminosity estimate and could probe the existence of a circum-secondary disk with more certainty than our observations. High resolution spectroscopy from HiRISE \citep{Vigan2023} could assess whether the companion is truly enhanced in metallicity. Epoch astrometry from \textit{Gaia} DR4 should improve the precision of the companion's dynamical mass estimate further, and enable a comprehensive benchmarking of multiple evolutionary models at very young ages. 

% \section*{Acknowledgements}
\begin{acknowledgements}
\par Thanks to our night astronomers for these observations, Thomas Rivinius, Claudia Paladini, Aaron Labdon, Abigail Frost, and Xavier Haubois. Thanks also to our telescope operators for these observations, Marcelo Lopez, Rodrigo Palominos, and Israel Blanchard. Special thanks to the Paranal and ESO staff for their support. We thank the anonymous reviewer for their constructive, encouraging report.
\par This work is based on observations collected at the European Southern Observatory under ESO programmes 109.237J.001 and 1104.C-0651.
\par S. L. acknowledges the support of the French Agence Nationale de la Recherche (ANR), under grant ANR-21-CE31-0017 (project ExoVLTI). 
\par J.J.W., A.C., and S.B. are supported by NASA XRP Grant 80NSSC23K0280.
\par G-D.M acknowledges the support of the DFG priority program SPP 1992 ``Exploring the Diversity of Extrasolar Planets'' (MA~9185/1) and from the Swiss National Science Foundation under grant 200021\_204847 ``PlanetsInTime''. Parts of this work have been carried out within the framework of the NCCR PlanetS supported by the Swiss National Science Foundation.
\par R.G.L. acknowledges the support of Science Foundation  Ireland under Grant No. 18/SIRG/5597.

\par This research has made use of the VizieR catalogue access tool, CDS, Strasbourg, France (DOI: 10.26093/cds/vizier).
\par This research has made use of the Jean-Marie Mariotti Center \texttt{Aspro} service.
\par This publication makes use of data products from the Two Micron All Sky Survey, which is a joint project of the University of Massachusetts and the Infrared Processing and Analysis Center/California Institute of Technology, funded by the National Aeronautics and Space Administration and the National Science Foundation. 
\par This work has made use of data from the European Space Agency (ESA) mission {\it Gaia} (\url{https://www.cosmos.esa.int/gaia}), processed by the {\it Gaia} Data Processing and Analysis Consortium (DPAC, \url{https://www.cosmos.esa.int/web/gaia/dpac/consortium}). Funding for the DPAC has been provided by national institutions, in particular the institutions participating in the {\it Gaia} Multilateral Agreement.

\par WOB acknowledges that the Johns Hopkins University occupies the unceded land of the Piscataway People, and acknowledges the Piscataway community, their elders both past and present, as well as future generations.
\par WOB acknowledges their cat, Morgoth, for her ``encouragement."

\end{acknowledgements}

\appendix

\section{Posterior distributions}
\par
\begin{figure}
    \centering
    \includegraphics[width=\textwidth]{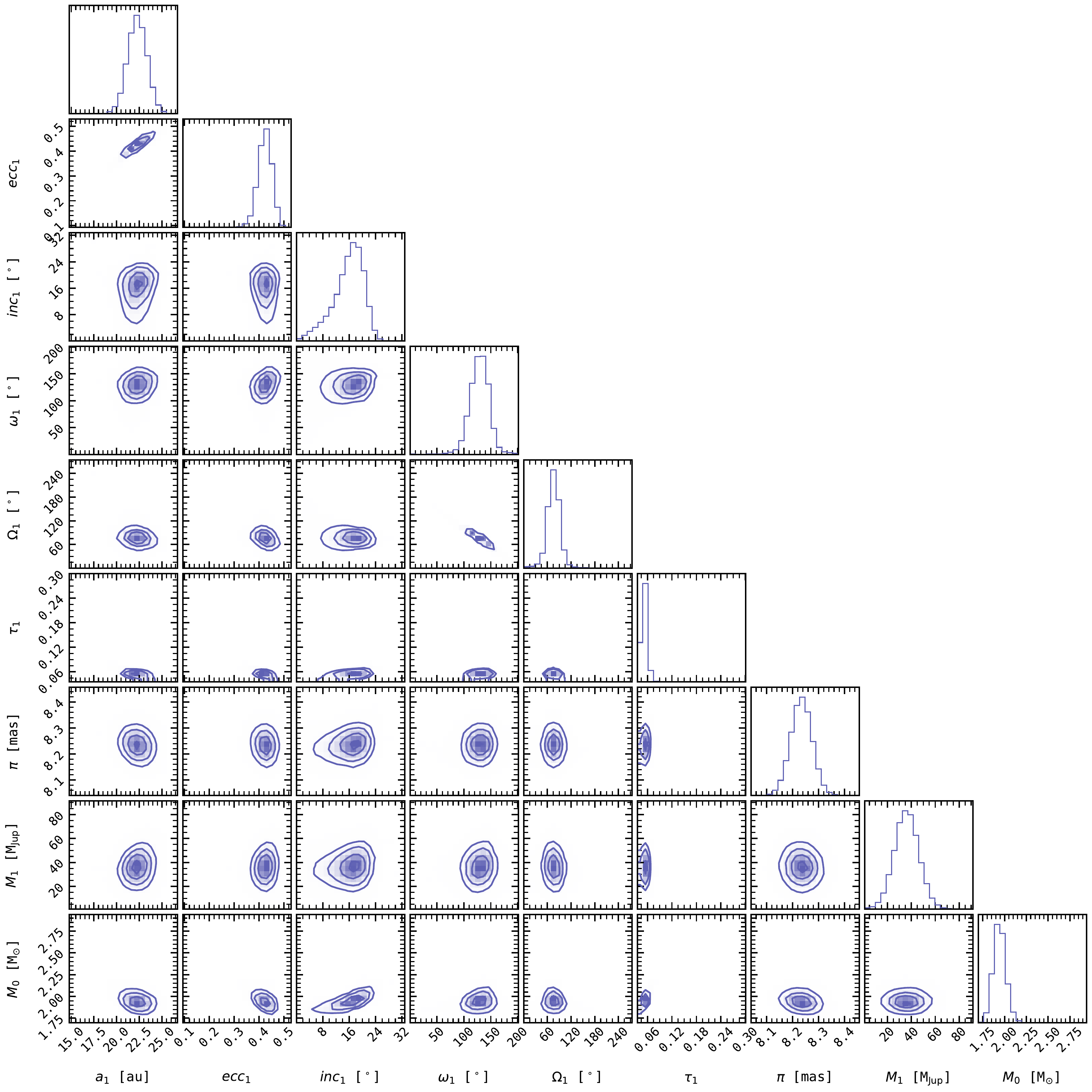}
    \caption{Posterior distribution of orbital elements for the orbital analysis presented in \S\ref{subsec:orbit}, and visualized in Figure \ref{fig:orbit}. The median and $1\,\sigma$ confidence intervals from the marginalized 1-D histograms show here, for each parameter, are recorded in Table \ref{tab:orbit}. Orbital parameters with respect to HD~136164~Ab are denoted with subscript ``1," and parameters with respect to HD~136164~A with subscript ``0." The distribution of orbital elements is unimodal, with no obvious degeneracies present between parameters, thanks to the combination of precise relative astrometry from GRAVITY, and long time baseline absolute astrometry from the HGCA considered in the fit.}
    \label{fig:orbit-post-full}
\end{figure}

\begin{figure}
    \centering
    \includegraphics[width=\textwidth]{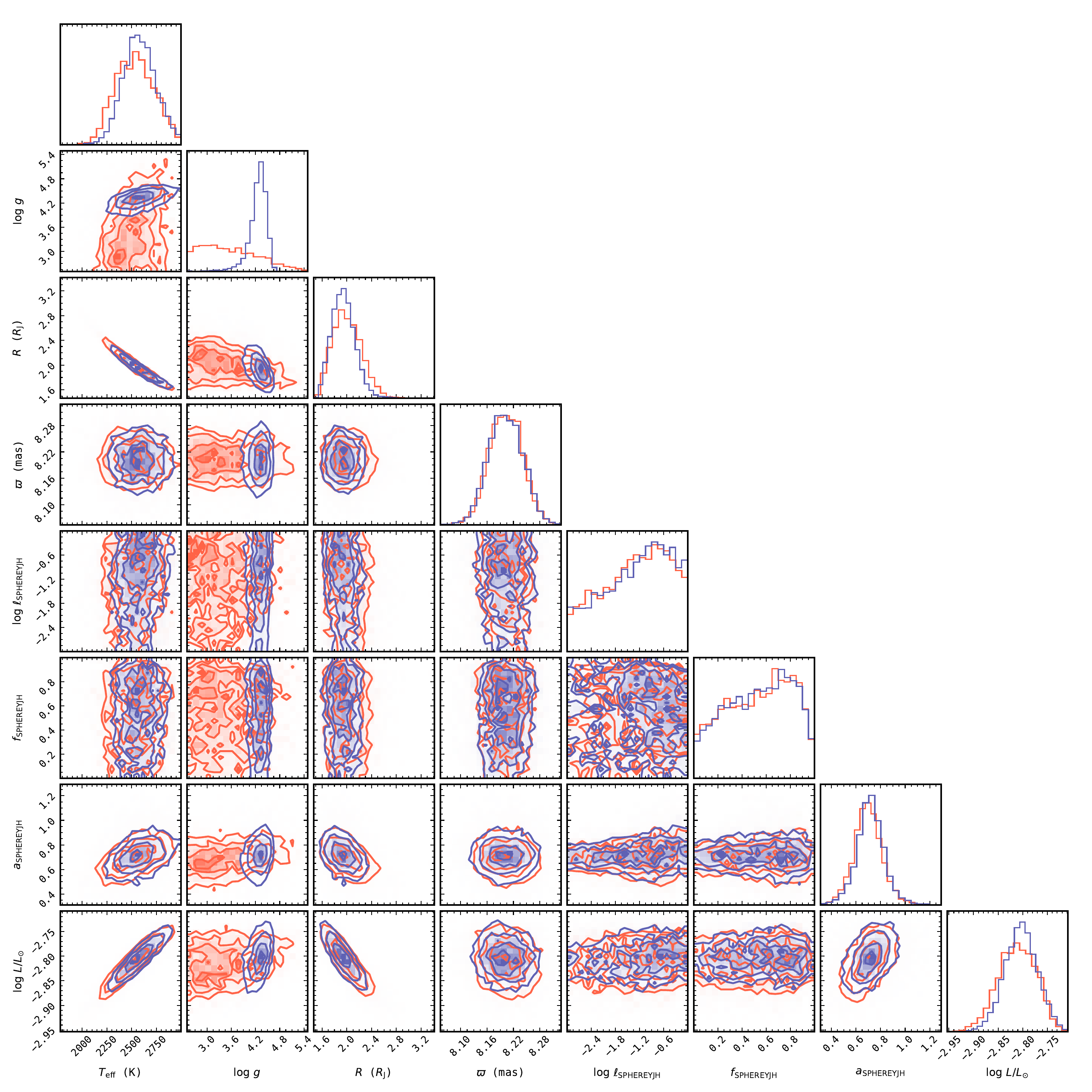}
    \caption{The posterior distribution of atmospheric model parameters from our \texttt{ATMO} spectral fit, in \S\ref{subsec:spectrum}. The contours of the posterior from the fit with a uniform mass prior are plotted in red, and the contours of the posterior from the fit with a dynamical mass prior are plotted in blue. The major takeaway is that $\log(g)$ and mass remain relatively unconstrained in the uniform prior fit, but the other parameters remain effectively the same between the two.}
    \label{fig:atmo-posterior-compare}
\end{figure}

\begin{figure}
    \centering
    \includegraphics[width=\textwidth]{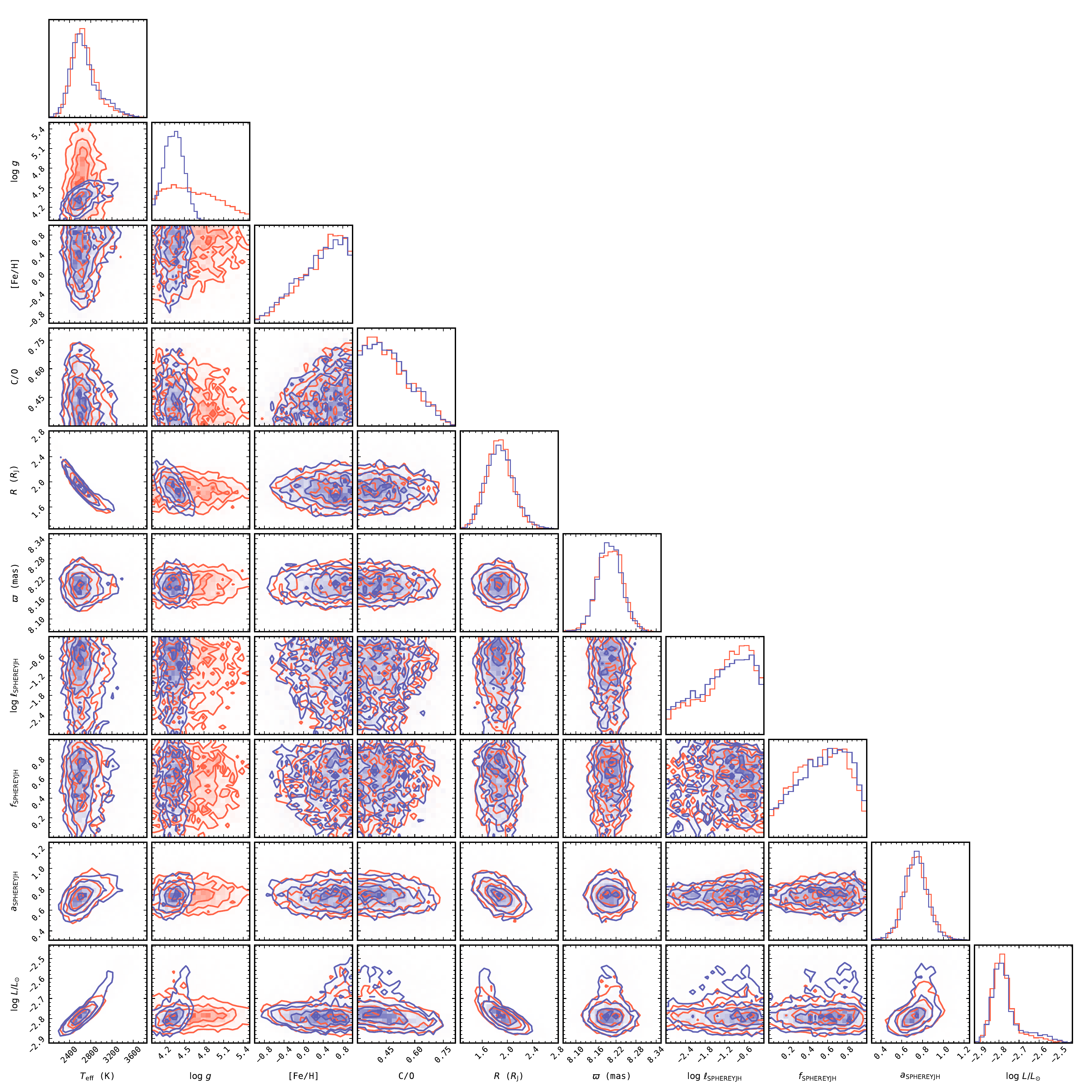}
    \caption{The posterior distribution of atmospheric model parameters from our \texttt{SPHINX} spectral fit, in \S\ref{subsec:spectrum}. The contours of the posterior from the fit with a uniform mass prior are plotted in red, and the contours of the posterior from the fit with a dynamical mass prior are plotted in blue. The major takeaway is that, with a slightly enhanced metallicity and sub-solar C/O, the sampler is no-longer driven towards implausibly low $\log(g)$.}
    \label{fig:sphinx-posterior-compare}
\end{figure}

 This appendix includes corner plots representing the posterior distributions for the two model fits in this work. The posterior distribution of orbital parameters from our \texttt{orbitize!} model fit are shown in Figure \ref{fig:orbit-post-full}. The posterior distribution on atmospheric parameters from our \texttt{ATMO} spectral fit using \texttt{species} is shown in Figure \ref{fig:atmo-posterior-compare}. The posterior distribution on atmospheric parameters from our \texttt{SPHINX} spectral fit using \texttt{species} is shown in Figure \ref{fig:sphinx-posterior-compare}.

\bibliography{hd136164ab_gravity}{}

\begin{thebibliography}{}
\providecommand\natexlab[1]{#1}
\providecommand\JournalTitle[1]{#1}

\bibitem[{{Allard} {et~al.}(2011){Allard}, {Homeier}, \&
  {Freytag}}]{Allard2011}
{Allard}, F., {Homeier}, D., \& {Freytag}, B. 2011,
  \href{http://dx.doi.org/10.48550/arXiv.1011.5405}{in Astronomical Society of
  the Pacific Conference Series, Vol. 448, 16th Cambridge Workshop on Cool
  Stars, Stellar Systems, and the Sun, ed. C.~{Johns-Krull}, M.~K. {Browning},
  \& A.~A. {West}}, 91

\bibitem[{{Astropy Collaboration} {et~al.}(2013){Astropy Collaboration},
  {Robitaille}, {Tollerud}, {Greenfield}, {Droettboom}, {Bray}, {Aldcroft},
  {Davis}, {Ginsburg}, {Price-Whelan}, {Kerzendorf}, {Conley}, {Crighton},
  {Barbary}, {Muna}, {Ferguson}, {Grollier}, {Parikh}, {Nair}, {Unther},
  {Deil}, {Woillez}, {Conseil}, {Kramer}, {Turner}, {Singer}, {Fox}, {Weaver},
  {Zabalza}, {Edwards}, {Azalee Bostroem}, {Burke}, {Casey}, {Crawford},
  {Dencheva}, {Ely}, {Jenness}, {Labrie}, {Lim}, {Pierfederici}, {Pontzen},
  {Ptak}, {Refsdal}, {Servillat}, \& {Streicher}}]{astropy:2013}
{Astropy Collaboration}, {Robitaille}, T.~P., {Tollerud}, E.~J., {et~al.} 2013,
  \href{http://dx.doi.org/10.1051/0004-6361/201322068}{\JournalTitle{\aap},
  558, A33}

\bibitem[{{Astropy Collaboration} {et~al.}(2018){Astropy Collaboration},
  {Price-Whelan}, {Sip{\H{o}}cz}, {G{\"u}nther}, {Lim}, {Crawford}, {Conseil},
  {Shupe}, {Craig}, {Dencheva}, {Ginsburg}, {Vand erPlas}, {Bradley},
  {P{\'e}rez-Su{\'a}rez}, {de Val-Borro}, {Aldcroft}, {Cruz}, {Robitaille},
  {Tollerud}, {Ardelean}, {Babej}, {Bach}, {Bachetti}, {Bakanov}, {Bamford},
  {Barentsen}, {Barmby}, {Baumbach}, {Berry}, {Biscani}, {Boquien}, {Bostroem},
  {Bouma}, {Brammer}, {Bray}, {Breytenbach}, {Buddelmeijer}, {Burke},
  {Calderone}, {Cano Rodr{\'\i}guez}, {Cara}, {Cardoso}, {Cheedella}, {Copin},
  {Corrales}, {Crichton}, {D'Avella}, {Deil}, {Depagne}, {Dietrich}, {Donath},
  {Droettboom}, {Earl}, {Erben}, {Fabbro}, {Ferreira}, {Finethy}, {Fox},
  {Garrison}, {Gibbons}, {Goldstein}, {Gommers}, {Greco}, {Greenfield},
  {Groener}, {Grollier}, {Hagen}, {Hirst}, {Homeier}, {Horton}, {Hosseinzadeh},
  {Hu}, {Hunkeler}, {Ivezi{\'c}}, {Jain}, {Jenness}, {Kanarek}, {Kendrew},
  {Kern}, {Kerzendorf}, {Khvalko}, {King}, {Kirkby}, {Kulkarni}, {Kumar},
  {Lee}, {Lenz}, {Littlefair}, {Ma}, {Macleod}, {Mastropietro}, {McCully},
  {Montagnac}, {Morris}, {Mueller}, {Mumford}, {Muna}, {Murphy}, {Nelson},
  {Nguyen}, {Ninan}, {N{\"o}the}, {Ogaz}, {Oh}, {Parejko}, {Parley}, {Pascual},
  {Patil}, {Patil}, {Plunkett}, {Prochaska}, {Rastogi}, {Reddy Janga},
  {Sabater}, {Sakurikar}, {Seifert}, {Sherbert}, {Sherwood-Taylor}, {Shih},
  {Sick}, {Silbiger}, {Singanamalla}, {Singer}, {Sladen}, {Sooley},
  {Sornarajah}, {Streicher}, {Teuben}, {Thomas}, {Tremblay}, {Turner},
  {Terr{\'o}n}, {van Kerkwijk}, {de la Vega}, {Watkins}, {Weaver}, {Whitmore},
  {Woillez}, {Zabalza}, \& {Astropy Contributors}}]{astropy:2018}
{Astropy Collaboration}, {Price-Whelan}, A.~M., {Sip{\H{o}}cz}, B.~M., {et~al.}
  2018, \href{http://dx.doi.org/10.3847/1538-3881/aabc4f}{\JournalTitle{\aj},
  156, 123}

\bibitem[{{Astropy Collaboration} {et~al.}(2022){Astropy Collaboration},
  {Price-Whelan}, {Lim}, {Earl}, {Starkman}, {Bradley}, {Shupe}, {Patil},
  {Corrales}, {Brasseur}, {N{"o}the}, {Donath}, {Tollerud}, {Morris},
  {Ginsburg}, {Vaher}, {Weaver}, {Tocknell}, {Jamieson}, {van Kerkwijk},
  {Robitaille}, {Merry}, {Bachetti}, {G{"u}nther}, {Aldcroft},
  {Alvarado-Montes}, {Archibald}, {B{'o}di}, {Bapat}, {Barentsen}, {Baz{'a}n},
  {Biswas}, {Boquien}, {Burke}, {Cara}, {Cara}, {Conroy}, {Conseil}, {Craig},
  {Cross}, {Cruz}, {D'Eugenio}, {Dencheva}, {Devillepoix}, {Dietrich},
  {Eigenbrot}, {Erben}, {Ferreira}, {Foreman-Mackey}, {Fox}, {Freij}, {Garg},
  {Geda}, {Glattly}, {Gondhalekar}, {Gordon}, {Grant}, {Greenfield}, {Groener},
  {Guest}, {Gurovich}, {Handberg}, {Hart}, {Hatfield-Dodds}, {Homeier},
  {Hosseinzadeh}, {Jenness}, {Jones}, {Joseph}, {Kalmbach}, {Karamehmetoglu},
  {Ka{l}uszy{'n}ski}, {Kelley}, {Kern}, {Kerzendorf}, {Koch}, {Kulumani},
  {Lee}, {Ly}, {Ma}, {MacBride}, {Maljaars}, {Muna}, {Murphy}, {Norman},
  {O'Steen}, {Oman}, {Pacifici}, {Pascual}, {Pascual-Granado}, {Patil},
  {Perren}, {Pickering}, {Rastogi}, {Roulston}, {Ryan}, {Rykoff}, {Sabater},
  {Sakurikar}, {Salgado}, {Sanghi}, {Saunders}, {Savchenko}, {Schwardt},
  {Seifert-Eckert}, {Shih}, {Jain}, {Shukla}, {Sick}, {Simpson},
  {Singanamalla}, {Singer}, {Singhal}, {Sinha}, {Sip{H{o}}cz}, {Spitler},
  {Stansby}, {Streicher}, {{{S}}umak}, {Swinbank}, {Taranu}, {Tewary},
  {Tremblay}, {Val-Borro}, {Van Kooten}, {Vasovi{'c}}, {Verma}, {de Miranda
  Cardoso}, {Williams}, {Wilson}, {Winkel}, {Wood-Vasey}, {Xue}, {Yoachim},
  {Zhang}, {Zonca}, \& {Astropy Project Contributors}}]{astropy:2022}
{Astropy Collaboration}, {Price-Whelan}, A.~M., {Lim}, P.~L., {et~al.} 2022,
  \href{http://dx.doi.org/10.3847/1538-4357/ac7c74}{\JournalTitle{apj}, 935,
  167}

\bibitem[{{Balmer} {et~al.}(2023){Balmer}, {Pueyo}, {Stolker}, {Reggiani},
  {Lacour}, {Maire}, {Molli{\`e}re}, {Nowak}, {Sing}, {Pourr{\'e}}, {Blunt},
  {Wang}, {Rickman}, {Henning}, {Ward-Duong}, {Abuter}, {Amorim},
  {Asensio-Torres}, {Benisty}, {Berger}, {Beust}, {Boccaletti}, {Bohn},
  {Bonnefoy}, {Bonnet}, {Bourdarot}, {Brandner}, {Cantalloube}, {Caselli},
  {Charnay}, {Chauvin}, {Chavez}, {Choquet}, {Christiaens}, {Cl{\'e}net},
  {Coud{\'e} du Foresto}, {Cridland}, {Dembet}, {Drescher}, {Duvert}, {Eckart},
  {Eisenhauer}, {Feuchtgruber}, {Garcia}, {Garcia Lopez}, {Gardner}, {Gendron},
  {Genzel}, {Gillessen}, {Girard}, {Haubois}, {Hei{\ss}el}, {Hinkley},
  {Hippler}, {Horrobin}, {Houll{\'e}}, {Hubert}, {Jocou}, {Kammerer},
  {Keppler}, {Kervella}, {Kreidberg}, {Lagrange}, {Lapeyr{\`e}re}, {Le
  Bouquin}, {L{\'e}na}, {Lutz}, {Mang}, {Marleau}, {M{\'e}rand}, {Monnier},
  {Mordasini}, {Mouillet}, {Nasedkin}, {Ott}, {Otten}, {Paladini}, {Paumard},
  {Perraut}, {Perrin}, {Pfuhl}, {Rameau}, {Rodet}, {Rustamkulov}, {Shangguan},
  {Shimizu}, {Straubmeier}, {Sturm}, {Tacconi}, {van Dishoeck}, {Vigan},
  {Vincent}, {Widmann}, {Wieprecht}, {Wiezorrek}, {Winterhalder}, {Woillez},
  {Yazici}, \& {Young}}]{Balmer2023}
{Balmer}, W.~O., {Pueyo}, L., {Stolker}, T., {et~al.} 2023,
  \href{http://dx.doi.org/10.48550/arXiv.2309.04403}{\JournalTitle{arXiv
  e-prints}, arXiv:2309.04403}

\bibitem[{{Baraffe} {et~al.}(2003){Baraffe}, {Chabrier}, {Allard}, \&
  {Hauschildt}}]{Baraffe2003}
{Baraffe}, I., {Chabrier}, G., {Allard}, F., \& {Hauschildt}, P. 2003, in Brown
  Dwarfs, ed. E.~{Mart{\'\i}n}, Vol. 211, 41

\bibitem[{{Bate}(2009)}]{Bate2009}
{Bate}, M.~R. 2009,
  \href{http://dx.doi.org/10.1111/j.1365-2966.2008.14106.x}{\JournalTitle{\mnras},
  392, 590}

\bibitem[{{Bate}(2012)}]{Bate2012}
{Bate}, M.~R. 2012,
  \href{http://dx.doi.org/10.1111/j.1365-2966.2011.19955.x}{\JournalTitle{\mnras},
  419, 3115}

\bibitem[{{Beuzit} {et~al.}(2019){Beuzit}, {Vigan}, {Mouillet}, {Dohlen},
  {Gratton}, {Boccaletti}, {Sauvage}, {Schmid}, {Langlois}, {Petit},
  {Baruffolo}, {Feldt}, {Milli}, {Wahhaj}, {Abe}, {Anselmi}, {Antichi},
  {Barette}, {Baudrand}, {Baudoz}, {Bazzon}, {Bernardi}, {Blanchard}, {Brast},
  {Bruno}, {Buey}, {Carbillet}, {Carle}, {Cascone}, {Chapron}, {Charton},
  {Chauvin}, {Claudi}, {Costille}, {De Caprio}, {de Boer}, {Delboulb{\'e}},
  {Desidera}, {Dominik}, {Downing}, {Dupuis}, {Fabron}, {Fantinel}, {Farisato},
  {Feautrier}, {Fedrigo}, {Fusco}, {Gigan}, {Ginski}, {Girard}, {Giro},
  {Gisler}, {Gluck}, {Gry}, {Henning}, {Hubin}, {Hugot}, {Incorvaia}, {Jaquet},
  {Kasper}, {Lagadec}, {Lagrange}, {Le Coroller}, {Le Mignant}, {Le Ruyet},
  {Lessio}, {Lizon}, {Llored}, {Lundin}, {Madec}, {Magnard}, {Marteaud},
  {Martinez}, {Maurel}, {M{\'e}nard}, {Mesa}, {M{\"o}ller-Nilsson}, {Moulin},
  {Moutou}, {Orign{\'e}}, {Parisot}, {Pavlov}, {Perret}, {Pragt}, {Puget},
  {Rabou}, {Ramos}, {Reess}, {Rigal}, {Rochat}, {Roelfsema}, {Rousset}, {Roux},
  {Saisse}, {Salasnich}, {Santambrogio}, {Scuderi}, {Segransan}, {Sevin},
  {Siebenmorgen}, {Soenke}, {Stadler}, {Suarez}, {Tiph{\`e}ne}, {Turatto},
  {Udry}, {Vakili}, {Waters}, {Weber}, {Wildi}, {Zins}, \&
  {Zurlo}}]{Beuzit2019}
{Beuzit}, J.~L., {Vigan}, A., {Mouillet}, D., {et~al.} 2019,
  \href{http://dx.doi.org/10.1051/0004-6361/201935251}{\JournalTitle{\aap},
  631, A155}

\bibitem[{{Biller} {et~al.}(2010){Biller}, {Liu}, {Wahhaj}, {Nielsen}, {Close},
  {Dupuy}, {Hayward}, {Burrows}, {Chun}, {Ftaclas}, {Clarke}, {Hartung},
  {Males}, {Reid}, {Shkolnik}, {Skemer}, {Tecza}, {Thatte}, {Alencar},
  {Artymowicz}, {Boss}, {de Gouveia Dal Pino}, {Gregorio-Hetem}, {Ida},
  {Kuchner}, {Lin}, \& {Toomey}}]{Biller2010}
{Biller}, B.~A., {Liu}, M.~C., {Wahhaj}, Z., {et~al.} 2010,
  \href{http://dx.doi.org/10.1088/2041-8205/720/1/L82}{\JournalTitle{\apjl},
  720, L82}

\bibitem[{{Blunt} {et~al.}(2020){Blunt}, {Wang}, {Angelo}, {Ngo}, {Cody}, {De
  Rosa}, {Graham}, {Hirsch}, {Nagpal}, {Nielsen}, {Pearce}, {Rice}, \&
  {Tejada}}]{Blunt2020}
{Blunt}, S., {Wang}, J.~J., {Angelo}, I., {et~al.} 2020,
  \href{http://dx.doi.org/10.3847/1538-3881/ab6663}{\JournalTitle{\aj}, 159,
  89}

\bibitem[{{Blunt} {et~al.}(2023){Blunt}, {Balmer}, {Wang}, {Lacour}, {Petrus},
  {Bourdarot}, {Kammerer}, {Pourr{\'e}}, {Rickman}, {Shangguan},
  {Winterhalder}, {Abuter}, {Amorim}, {Asensio-Torres}, {Benisty}, {Berger},
  {Beust}, {Boccaletti}, {Bohn}, {Bonnefoy}, {Bonnet}, {Brandner},
  {Cantalloube}, {Caselli}, {Charnay}, {Chauvin}, {Chavez}, {Choquet},
  {Christiaens}, {Cl{\'e}net}, {Du Foresto}, {Cridland}, {Dembet}, {Drescher},
  {Duvert}, {Eckart}, {Eisenhauer}, {Feuchtgruber}, {Garcia}, {Garcia Lopez},
  {Gendron}, {Genzel}, {Gillessen}, {Girard}, {Haubois}, {Hei{\ss}el},
  {Henning}, {Hinkley}, {Hippler}, {Horrobin}, {Houll{\'e}}, {Hubert}, {Jocou},
  {Keppler}, {Kervella}, {Kreidberg}, {Lagrange}, {Lapeyr{\`e}re}, {Le
  Bouquin}, {L{\'e}na}, {Lutz}, {Maire}, {Mang}, {Marleau}, {M{\'e}rand},
  {Molli{\`e}re}, {Monnier}, {Mordasini}, {Mouillet}, {Nasedkin}, {Nowak},
  {Ott}, {Otten}, {Paladini}, {Paumard}, {Perraut}, {Perrin}, {Pfuhl}, {Pueyo},
  {Rameau}, {Rodet}, {Rustamkulov}, {Shimizu}, {Sing}, {Stolker},
  {Straubmeier}, {Sturm}, {Tacconi}, {van Dishoeck}, {Vigan}, {Vincent},
  {Ward-Duong}, {Widmann}, {Wieprecht}, {Wiezorrek}, {Woillez}, {Yazici},
  {Young}, \& {Exogravity Collaboration}}]{Blunt2023}
{Blunt}, S., {Balmer}, W.~O., {Wang}, J.~J., {et~al.} 2023,
  \href{http://dx.doi.org/10.3847/1538-3881/ad06b7}{\JournalTitle{\aj}, 166,
  257}

\bibitem[{{Bochanski} {et~al.}(2018){Bochanski}, {Faherty}, {Gagn{\'e}},
  {Nelson}, {Coker}, {Smithka}, {Desir}, \& {Vasquez}}]{Bochanski2018}
{Bochanski}, J.~J., {Faherty}, J.~K., {Gagn{\'e}}, J., {et~al.} 2018,
  \href{http://dx.doi.org/10.3847/1538-3881/aaaebe}{\JournalTitle{\aj}, 155,
  149}

\bibitem[{{Bodenheimer} {et~al.}(2013){Bodenheimer}, {D'Angelo}, {Lissauer},
  {Fortney}, \& {Saumon}}]{Bodenheimer2013}
{Bodenheimer}, P., {D'Angelo}, G., {Lissauer}, J.~J., {Fortney}, J.~J., \&
  {Saumon}, D. 2013,
  \href{http://dx.doi.org/10.1088/0004-637X/770/2/120}{\JournalTitle{\apj},
  770, 120}

\bibitem[{{Bohn} {et~al.}(2022){Bohn}, {Ginski}, {Kenworthy}, {Mamajek},
  {Meshkat}, {Pecaut}, {Reggiani}, {Seay}, {Brown}, {Cugno}, {Henning},
  {Launhardt}, {Quirrenbach}, {Rickman}, \& {S{\'e}gransan}}]{Bohn2022}
{Bohn}, A.~J., {Ginski}, C., {Kenworthy}, M.~A., {et~al.} 2022,
  \href{http://dx.doi.org/10.1051/0004-6361/202039917}{\JournalTitle{\aap},
  657, A53}

\bibitem[{{Bonavita} {et~al.}(2022){Bonavita}, {Fontanive}, {Gratton},
  {Mu{\v{z}}i{\'c}}, {Desidera}, {Mesa}, {Biller}, {Scholz}, {Sozzetti}, \&
  {Squicciarini}}]{Bonavita2022}
{Bonavita}, M., {Fontanive}, C., {Gratton}, R., {et~al.} 2022,
  \href{http://dx.doi.org/10.1093/mnras/stac1250}{\JournalTitle{\mnras}, 513,
  5588}

\bibitem[{{Boss}(1997)}]{Boss1997}
{Boss}, A.~P. 1997,
  \href{http://dx.doi.org/10.1126/science.276.5320.1836}{\JournalTitle{Science},
  276, 1836}

\bibitem[{{Bowler} {et~al.}(2020){Bowler}, {Blunt}, \& {Nielsen}}]{Bowler2020}
{Bowler}, B.~P., {Blunt}, S.~C., \& {Nielsen}, E.~L. 2020,
  \href{http://dx.doi.org/10.3847/1538-3881/ab5b11}{\JournalTitle{\aj}, 159,
  63}

\bibitem[{{Boyd} \& {Whitworth}(2005)}]{Boyd2005}
{Boyd}, D.~F.~A., \& {Whitworth}, A.~P. 2005,
  \href{http://dx.doi.org/10.1051/0004-6361:20041703}{\JournalTitle{\aap}, 430,
  1059}

\bibitem[{{Brandt} {et~al.}(2021){Brandt}, {Dupuy}, {Li}, {Chen}, {Brandt},
  {Wong}, {Currie}, {Bowler}, {Liu}, {Best}, \& {Phillips}}]{mBrandt2021}
{Brandt}, G.~M., {Dupuy}, T.~J., {Li}, Y., {et~al.} 2021,
  \href{http://dx.doi.org/10.3847/1538-3881/ac273e}{\JournalTitle{\aj}, 162,
  301}

\bibitem[{{Brandt}(2021)}]{Brandt2021}
{Brandt}, T.~D. 2021,
  \href{http://dx.doi.org/10.3847/1538-4365/abf93c}{\JournalTitle{\apjs}, 254,
  42}

\bibitem[{{Brandt} {et~al.}(2019){Brandt}, {Dupuy}, \& {Bowler}}]{Brandt2019}
{Brandt}, T.~D., {Dupuy}, T.~J., \& {Bowler}, B.~P. 2019,
  \href{http://dx.doi.org/10.3847/1538-3881/ab04a8}{\JournalTitle{\aj}, 158,
  140}

\bibitem[{{Buchner} {et~al.}(2014){Buchner}, {Georgakakis}, {Nandra}, {Hsu},
  {Rangel}, {Brightman}, {Merloni}, {Salvato}, {Donley}, \&
  {Kocevski}}]{Buchner2014}
{Buchner}, J., {Georgakakis}, A., {Nandra}, K., {et~al.} 2014,
  \href{http://dx.doi.org/10.1051/0004-6361/201322971}{\JournalTitle{\aap},
  564, A125}

\bibitem[{{Carnall}(2017)}]{Carnall2017}
{Carnall}, A.~C. 2017,
  \href{http://dx.doi.org/10.48550/arXiv.1705.05165}{\JournalTitle{arXiv
  e-prints}, arXiv:1705.05165}

\bibitem[{{Chabrier}(2003)}]{Chabrier2003}
{Chabrier}, G. 2003,
  \href{http://dx.doi.org/10.1086/376392}{\JournalTitle{\pasp}, 115, 763}

\bibitem[{{Chabrier} \& {Baraffe}(2000)}]{Chabrier2000}
{Chabrier}, G., \& {Baraffe}, I. 2000,
  \href{http://dx.doi.org/10.1146/annurev.astro.38.1.337}{\JournalTitle{\araa},
  38, 337}

\bibitem[{{Chauvin} {et~al.}(2017){Chauvin}, {Desidera}, {Lagrange}, {Vigan},
  {Gratton}, {Langlois}, {Bonnefoy}, {Beuzit}, {Feldt}, {Mouillet}, {Meyer},
  {Cheetham}, {Biller}, {Boccaletti}, {D'Orazi}, {Galicher}, {Hagelberg},
  {Maire}, {Mesa}, {Olofsson}, {Samland}, {Schmidt}, {Sissa}, {Bonavita},
  {Charnay}, {Cudel}, {Daemgen}, {Delorme}, {Janin-Potiron}, {Janson},
  {Keppler}, {Le Coroller}, {Ligi}, {Marleau}, {Messina}, {Molli{\`e}re},
  {Mordasini}, {M{\"u}ller}, {Peretti}, {Perrot}, {Rodet}, {Rouan}, {Zurlo},
  {Dominik}, {Henning}, {Menard}, {Schmid}, {Turatto}, {Udry}, {Vakili}, {Abe},
  {Antichi}, {Baruffolo}, {Baudoz}, {Baudrand}, {Blanchard}, {Bazzon}, {Buey},
  {Carbillet}, {Carle}, {Charton}, {Cascone}, {Claudi}, {Costille}, {Deboulbe},
  {De Caprio}, {Dohlen}, {Fantinel}, {Feautrier}, {Fusco}, {Gigan}, {Giro},
  {Gisler}, {Gluck}, {Hubin}, {Hugot}, {Jaquet}, {Kasper}, {Madec}, {Magnard},
  {Martinez}, {Maurel}, {Le Mignant}, {M{\"o}ller-Nilsson}, {Llored}, {Moulin},
  {Orign{\'e}}, {Pavlov}, {Perret}, {Petit}, {Pragt}, {Puget}, {Rabou},
  {Ramos}, {Rigal}, {Rochat}, {Roelfsema}, {Rousset}, {Roux}, {Salasnich},
  {Sauvage}, {Sevin}, {Soenke}, {Stadler}, {Suarez}, {Weber}, {Wildi},
  {Antoniucci}, {Augereau}, {Baudino}, {Brandner}, {Engler}, {Girard}, {Gry},
  {Kral}, {Kopytova}, {Lagadec}, {Milli}, {Moutou}, {Schlieder},
  {Szul{\'a}gyi}, {Thalmann}, \& {Wahhaj}}]{Chauvin2017}
{Chauvin}, G., {Desidera}, S., {Lagrange}, A.~M., {et~al.} 2017,
  \href{http://dx.doi.org/10.1051/0004-6361/201731152}{\JournalTitle{\aap},
  605, L9}

\bibitem[{{Cheetham} {et~al.}(2018){Cheetham}, {Bonnefoy}, {Desidera},
  {Langlois}, {Vigan}, {Schmidt}, {Olofsson}, {Chauvin}, {Klahr}, {Gratton},
  {D'Orazi}, {Henning}, {Janson}, {Biller}, {Peretti}, {Hagelberg},
  {S{\'e}gransan}, {Udry}, {Mesa}, {Sissa}, {Kral}, {Schlieder}, {Maire},
  {Mordasini}, {Menard}, {Zurlo}, {Beuzit}, {Feldt}, {Mouillet}, {Meyer},
  {Lagrange}, {Boccaletti}, {Keppler}, {Kopytova}, {Ligi}, {Rouan}, {Le
  Coroller}, {Dominik}, {Lagadec}, {Turatto}, {Abe}, {Antichi}, {Baruffolo},
  {Baudoz}, {Blanchard}, {Buey}, {Carbillet}, {Carle}, {Cascone}, {Claudi},
  {Costille}, {Delboulb{\'e}}, {De Caprio}, {Dohlen}, {Fantinel}, {Feautrier},
  {Fusco}, {Giro}, {Gluck}, {Hubin}, {Hugot}, {Jaquet}, {Kasper}, {Llored},
  {Madec}, {Magnard}, {Martinez}, {Maurel}, {Le Mignant}, {M{\"o}ller-Nilsson},
  {Moulin}, {Orign{\'e}}, {Pavlov}, {Perret}, {Petit}, {Pragt}, {Puget},
  {Rabou}, {Ramos}, {Rigal}, {Rochat}, {Roelfsema}, {Rousset}, {Roux},
  {Salasnich}, {Sauvage}, {Sevin}, {Soenke}, {Stadler}, {Suarez}, {Weber}, \&
  {Wildi}}]{Cheetham2018}
{Cheetham}, A., {Bonnefoy}, M., {Desidera}, S., {et~al.} 2018,
  \href{http://dx.doi.org/10.1051/0004-6361/201832650}{\JournalTitle{\aap},
  615, A160}

\bibitem[{{Chomez} {et~al.}(2023){Chomez}, {Squicciarini}, {Lagrange},
  {Delorme}, {Viswanath}, {Janson}, {Flasseur}, {Chauvin}, {Langlois},
  {Rubini}, {Bergeon}, {Albert}, {Bonnefoy}, {Desidera}, {Engler}, {Gratton},
  {Henning}, {Mamajek}, {Marleau}, {Meyer}, {Reffert}, {Ringqvist}, \&
  {Samland}}]{Chomez2023}
{Chomez}, A., {Squicciarini}, V., {Lagrange}, A.~M., {et~al.} 2023,
  \href{http://dx.doi.org/10.1051/0004-6361/202347044}{\JournalTitle{\aap},
  676, L10}

\bibitem[{{Claudi} {et~al.}(2008){Claudi}, {Turatto}, {Gratton}, {Antichi},
  {Bonavita}, {Bruno}, {Cascone}, {De Caprio}, {Desidera}, {Giro}, {Mesa},
  {Scuderi}, {Dohlen}, {Beuzit}, \& {Puget}}]{Claudi2008}
{Claudi}, R.~U., {Turatto}, M., {Gratton}, R.~G., {et~al.} 2008,
  \href{http://dx.doi.org/10.1117/12.788366}{in Society of Photo-Optical
  Instrumentation Engineers (SPIE) Conference Series, Vol. 7014, Ground-based
  and Airborne Instrumentation for Astronomy II, ed. I.~S. {McLean} \& M.~M.
  {Casali}}, 70143E

\bibitem[{{Cutri} {et~al.}(2003){Cutri}, {Skrutskie}, {van Dyk}, {Beichman},
  {Carpenter}, {Chester}, {Cambresy}, {Evans}, {Fowler}, {Gizis}, {Howard},
  {Huchra}, {Jarrett}, {Kopan}, {Kirkpatrick}, {Light}, {Marsh}, {McCallon},
  {Schneider}, {Stiening}, {Sykes}, {Weinberg}, {Wheaton}, {Wheelock}, \&
  {Zacarias}}]{Cutri2003}
{Cutri}, R.~M., {Skrutskie}, M.~F., {van Dyk}, S., {et~al.} 2003,
  \JournalTitle{VizieR Online Data Catalog}, II/246

\bibitem[{{D'Angelo} {et~al.}(2010){D'Angelo}, {Durisen}, \&
  {Lissauer}}]{D'Angelo2010}
{D'Angelo}, G., {Durisen}, R.~H., \& {Lissauer}, J.~J. 2010,
  \href{http://dx.doi.org/10.48550/arXiv.1006.5486}{in Exoplanets, ed.
  S.~{Seager}}, 319

\bibitem[{{Delorme} {et~al.}(2017){Delorme}, {Schmidt}, {Bonnefoy}, {Desidera},
  {Ginski}, {Charnay}, {Lazzoni}, {Christiaens}, {Messina}, {D'Orazi}, {Milli},
  {Schlieder}, {Gratton}, {Rodet}, {Lagrange}, {Absil}, {Vigan}, {Galicher},
  {Hagelberg}, {Bonavita}, {Lavie}, {Zurlo}, {Olofsson}, {Boccaletti},
  {Cantalloube}, {Mouillet}, {Chauvin}, {Hambsch}, {Langlois}, {Udry},
  {Henning}, {Beuzit}, {Mordasini}, {Lucas}, {Marocco}, {Biller}, {Carson},
  {Cheetham}, {Covino}, {De Caprio}, {Delboulbe}, {Feldt}, {Girard}, {Hubin},
  {Maire}, {Pavlov}, {Petit}, {Rouan}, {Roelfsema}, \& {Wildi}}]{Delorme2017}
{Delorme}, P., {Schmidt}, T., {Bonnefoy}, M., {et~al.} 2017,
  \href{http://dx.doi.org/10.1051/0004-6361/201731145}{\JournalTitle{\aap},
  608, A79}

\bibitem[{{Dieterich} {et~al.}(2012){Dieterich}, {Henry}, {Golimowski},
  {Krist}, \& {Tanner}}]{Dieterich2012}
{Dieterich}, S.~B., {Henry}, T.~J., {Golimowski}, D.~A., {Krist}, J.~E., \&
  {Tanner}, A.~M. 2012,
  \href{http://dx.doi.org/10.1088/0004-6256/144/2/64}{\JournalTitle{\aj}, 144,
  64}

\bibitem[{{Dieterich} {et~al.}(2021){Dieterich}, {Simler}, {Henry}, \&
  {Jao}}]{Dieterich2021}
{Dieterich}, S.~B., {Simler}, A., {Henry}, T.~J., \& {Jao}, W.-C. 2021,
  \href{http://dx.doi.org/10.3847/1538-3881/abd2c2}{\JournalTitle{\aj}, 161,
  172}

\bibitem[{{Do {\'O}} {et~al.}(2023){Do {\'O}}, {O'Neil}, {Konopacky}, {Do},
  {Martinez}, {Ruffio}, \& {Ghez}}]{DoO2023}
{Do {\'O}}, C.~R., {O'Neil}, K.~K., {Konopacky}, Q.~M., {et~al.} 2023,
  \href{http://dx.doi.org/10.48550/arXiv.2306.04080}{\JournalTitle{arXiv
  e-prints}, arXiv:2306.04080}

\bibitem[{{Dohlen} {et~al.}(2008){Dohlen}, {Langlois}, {Saisse}, {Hill},
  {Origne}, {Jacquet}, {Fabron}, {Blanc}, {Llored}, {Carle}, {Moutou}, {Vigan},
  {Boccaletti}, {Carbillet}, {Mouillet}, \& {Beuzit}}]{Dohlen2008}
{Dohlen}, K., {Langlois}, M., {Saisse}, M., {et~al.} 2008,
  \href{http://dx.doi.org/10.1117/12.789786}{in Society of Photo-Optical
  Instrumentation Engineers (SPIE) Conference Series, Vol. 7014, Ground-based
  and Airborne Instrumentation for Astronomy II, ed. I.~S. {McLean} \& M.~M.
  {Casali}}, 70143L

\bibitem[{{Duch{\^e}ne} {et~al.}(2023){Duch{\^e}ne}, {Oon}, {De Rosa},
  {Kantorski}, {Coy}, {Wang}, {Thomas}, {Patience}, {Pueyo}, {Nielsen}, \&
  {Konopacky}}]{Duchene2023}
{Duch{\^e}ne}, G., {Oon}, J.~T., {De Rosa}, R.~J., {et~al.} 2023,
  \href{http://dx.doi.org/10.1093/mnras/stac3527}{\JournalTitle{\mnras}, 519,
  778}

\bibitem[{{Dupuy} \& {Liu}(2017)}]{Dupuy2017}
{Dupuy}, T.~J., \& {Liu}, M.~C. 2017,
  \href{http://dx.doi.org/10.3847/1538-4365/aa5e4c}{\JournalTitle{\apjs}, 231,
  15}

\bibitem[{{Emsenhuber} {et~al.}(2021{\natexlab{a}}){Emsenhuber}, {Mordasini},
  {Burn}, {Alibert}, {Benz}, \& {Asphaug}}]{Emsenhuber2021}
{Emsenhuber}, A., {Mordasini}, C., {Burn}, R., {et~al.} 2021{\natexlab{a}},
  \href{http://dx.doi.org/10.1051/0004-6361/202038553}{\JournalTitle{\aap},
  656, A69}

\bibitem[{{Emsenhuber} {et~al.}(2021{\natexlab{b}}){Emsenhuber}, {Mordasini},
  {Burn}, {Alibert}, {Benz}, \& {Asphaug}}]{Emsenhuber2021a}
{Emsenhuber}, A., {Mordasini}, C., {Burn}, R., {et~al.} 2021{\natexlab{b}},
  \href{http://dx.doi.org/10.1051/0004-6361/202038553}{\JournalTitle{\aap},
  656, A69}

\bibitem[{{Emsenhuber} {et~al.}(2021{\natexlab{c}}){Emsenhuber}, {Mordasini},
  {Burn}, {Alibert}, {Benz}, \& {Asphaug}}]{Emsenhuber2021b}
{Emsenhuber}, A., {Mordasini}, C., {Burn}, R., {et~al.} 2021{\natexlab{c}},
  \href{http://dx.doi.org/10.1051/0004-6361/202038863}{\JournalTitle{\aap},
  656, A70}

\bibitem[{{Feroz} \& {Hobson}(2008)}]{Feroz2008}
{Feroz}, F., \& {Hobson}, M.~P. 2008,
  \href{http://dx.doi.org/10.1111/j.1365-2966.2007.12353.x}{\JournalTitle{\mnras},
  384, 449}

\bibitem[{{Feroz} {et~al.}(2009){Feroz}, {Hobson}, \& {Bridges}}]{Feroz2009}
{Feroz}, F., {Hobson}, M.~P., \& {Bridges}, M. 2009,
  \href{http://dx.doi.org/10.1111/j.1365-2966.2009.14548.x}{\JournalTitle{\mnras},
  398, 1601}

\bibitem[{{Fontanive} {et~al.}(2018){Fontanive}, {Biller}, {Bonavita}, \&
  {Allers}}]{Fontanive2018}
{Fontanive}, C., {Biller}, B., {Bonavita}, M., \& {Allers}, K. 2018,
  \href{http://dx.doi.org/10.1093/mnras/sty1682}{\JournalTitle{\mnras}, 479,
  2702}

\bibitem[{{Fontanive} {et~al.}(2019){Fontanive}, {Mu{\v{z}}i{\'c}}, {},
  {Bonavita}, \& {Biller}}]{Fontanive2019}
{Fontanive}, C., {Mu{\v{z}}i{\'c}}, {}, K., {Bonavita}, M., \& {Biller}, B.
  2019, \href{http://dx.doi.org/10.1093/mnras/stz2587}{\JournalTitle{\mnras},
  490, 1120}

\bibitem[{{Foreman-Mackey} {et~al.}(2013){Foreman-Mackey}, {Hogg}, {Lang}, \&
  {Goodman}}]{Foreman-Mackey2013}
{Foreman-Mackey}, D., {Hogg}, D.~W., {Lang}, D., \& {Goodman}, J. 2013,
  \href{http://dx.doi.org/10.1086/670067}{\JournalTitle{\pasp}, 125, 306}

\bibitem[{{Forgan} {et~al.}(2015){Forgan}, {Parker}, \& {Rice}}]{Forgan2015}
{Forgan}, D., {Parker}, R.~J., \& {Rice}, K. 2015,
  \href{http://dx.doi.org/10.1093/mnras/stu2504}{\JournalTitle{\mnras}, 447,
  836}

\bibitem[{{Forgan} \& {Rice}(2013)}]{Forgan2013}
{Forgan}, D., \& {Rice}, K. 2013,
  \href{http://dx.doi.org/10.1093/mnras/stt672}{\JournalTitle{\mnras}, 432,
  3168}

\bibitem[{{Forgan} {et~al.}(2018){Forgan}, {Hall}, {Meru}, \&
  {Rice}}]{Forgan2018}
{Forgan}, D.~H., {Hall}, C., {Meru}, F., \& {Rice}, W.~K.~M. 2018,
  \href{http://dx.doi.org/10.1093/mnras/stx2870}{\JournalTitle{\mnras}, 474,
  5036}

\bibitem[{{Franson} \& {Bowler}(2023)}]{Franson2023}
{Franson}, K., \& {Bowler}, B.~P. 2023,
  \href{http://dx.doi.org/10.3847/1538-3881/acca18}{\JournalTitle{\aj}, 165,
  246}

\bibitem[{{Franson} {et~al.}(2022){Franson}, {Bowler}, {Brandt}, {Dupuy},
  {Tran}, {Brandt}, {Li}, \& {Kraus}}]{Franson2022}
{Franson}, K., {Bowler}, B.~P., {Brandt}, T.~D., {et~al.} 2022,
  \href{http://dx.doi.org/10.3847/1538-3881/ac35e8}{\JournalTitle{\aj}, 163,
  50}

\bibitem[{{Gagn{\'e}} {et~al.}(2018){Gagn{\'e}}, {Mamajek}, {Malo}, {Riedel},
  {Rodriguez}, {Lafreni{\`e}re}, {Faherty}, {Roy-Loubier}, {Pueyo}, {Robin}, \&
  {Doyon}}]{Gagne2018}
{Gagn{\'e}}, J., {Mamajek}, E.~E., {Malo}, L., {et~al.} 2018,
  \href{http://dx.doi.org/10.3847/1538-4357/aaae09}{\JournalTitle{\apj}, 856,
  23}

\bibitem[{{Gaia Collaboration} {et~al.}(2022){Gaia Collaboration}, {Vallenari},
  {Brown}, {Prusti}, {de Bruijne}, {Arenou}, {Babusiaux}, {Biermann},
  {Creevey}, {Ducourant}, \& et~al.}]{GaiaCollaboration2022}
{Gaia Collaboration}, {Vallenari}, A., {Brown}, A.~G.~A., {et~al.} 2022,
  \href{http://dx.doi.org/10.48550/arXiv.2208.00211}{\JournalTitle{arXiv
  e-prints}, arXiv:2208.00211}

\bibitem[{{Girardi} {et~al.}(2000){Girardi}, {Bressan}, {Bertelli}, \&
  {Chiosi}}]{Girardi2000}
{Girardi}, L., {Bressan}, A., {Bertelli}, G., \& {Chiosi}, C. 2000,
  \href{http://dx.doi.org/10.1051/aas:2000126}{\JournalTitle{\aaps}, 141, 371}

\bibitem[{{Grandjean} {et~al.}(2019){Grandjean}, {Lagrange}, {Beust}, {Rodet},
  {Milli}, {Rubini}, {Babusiaux}, {Meunier}, {Delorme}, {Aigrain}, {Zicher},
  {Bonnefoy}, {Biller}, {Baudino}, {Bonavita}, {Boccaletti}, {Cheetham},
  {Girard}, {Hagelberg}, {Janson}, {Lannier}, {Lazzoni}, {Ligi}, {Maire},
  {Mesa}, {Perrot}, {Rouan}, \& {Zurlo}}]{Grandjean2019}
{Grandjean}, A., {Lagrange}, A.~M., {Beust}, H., {et~al.} 2019,
  \href{http://dx.doi.org/10.1051/0004-6361/201935044}{\JournalTitle{\aap},
  627, L9}

\bibitem[{{Gravity Collaboration} {et~al.}(2017){Gravity Collaboration},
  {Abuter}, {Accardo}, {Amorim}, {Anugu}, {{\'A}vila}, {Azouaoui}, {Benisty},
  {Berger}, {Blind}, {Bonnet}, {Bourget}, {Brandner}, {Brast}, {Buron},
  {Burtscher}, {Cassaing}, {Chapron}, {Choquet}, {Cl{\'e}net}, {Collin},
  {Coud{\'e} Du Foresto}, {de Wit}, {de Zeeuw}, {Deen},
  {Delplancke-Str{\"o}bele}, {Dembet}, {Derie}, {Dexter}, {Duvert}, {Ebert},
  {Eckart}, {Eisenhauer}, {Esselborn}, {F{\'e}dou}, {Finger}, {Garcia}, {Garcia
  Dabo}, {Garcia Lopez}, {Gendron}, {Genzel}, {Gillessen}, {Gonte}, {Gordo},
  {Grould}, {Gr{\"o}zinger}, {Guieu}, {Haguenauer}, {Hans}, {Haubois}, {Haug},
  {Haussmann}, {Henning}, {Hippler}, {Horrobin}, {Huber}, {Hubert}, {Hubin},
  {Hummel}, {Jakob}, {Janssen}, {Jochum}, {Jocou}, {Kaufer}, {Kellner},
  {Kendrew}, {Kern}, {Kervella}, {Kiekebusch}, {Klein}, {Kok}, {Kolb}, {Kulas},
  {Lacour}, {Lapeyr{\`e}re}, {Lazareff}, {Le Bouquin}, {L{\`e}na}, {Lenzen},
  {L{\'e}v{\^e}que}, {Lippa}, {Magnard}, {Mehrgan}, {Mellein}, {M{\'e}rand},
  {Moreno-Ventas}, {Moulin}, {M{\"u}ller}, {M{\"u}ller}, {Neumann}, {Oberti},
  {Ott}, {Pallanca}, {Panduro}, {Pasquini}, {Paumard}, {Percheron}, {Perraut},
  {Perrin}, {Pfl{\"u}ger}, {Pfuhl}, {Phan Duc}, {Plewa}, {Popovic}, {Rabien},
  {Ram{\'\i}rez}, {Ramos}, {Rau}, {Riquelme}, {Rohloff}, {Rousset},
  {Sanchez-Bermudez}, {Scheithauer}, {Sch{\"o}ller}, {Schuhler}, {Spyromilio},
  {Straubmeier}, {Sturm}, {Suarez}, {Tristram}, {Ventura}, {Vincent},
  {Waisberg}, {Wank}, {Weber}, {Wieprecht}, {Wiest}, {Wiezorrek}, {Wittkowski},
  {Woillez}, {Wolff}, {Yazici}, {Ziegler}, \&
  {Zins}}]{GravityCollaboration2017}
{Gravity Collaboration}, {Abuter}, R., {Accardo}, M., {et~al.} 2017,
  \href{http://dx.doi.org/10.1051/0004-6361/201730838}{\JournalTitle{\aap},
  602, A94}

\bibitem[{{GRAVITY Collaboration} {et~al.}(2019){GRAVITY Collaboration},
  {Lacour}, {Nowak}, {Wang}, {Pfuhl}, {Eisenhauer}, {Abuter}, {Amorim},
  {Anugu}, {Benisty}, {Berger}, {Beust}, {Blind}, {Bonnefoy}, {Bonnet},
  {Bourget}, {Brandner}, {Buron}, {Collin}, {Charnay}, {Chapron}, {Cl{\'e}net},
  {Coud{\'e} Du Foresto}, {de Zeeuw}, {Deen}, {Dembet}, {Dexter}, {Duvert},
  {Eckart}, {F{\"o}rster Schreiber}, {F{\'e}dou}, {Garcia}, {Garcia Lopez},
  {Gao}, {Gendron}, {Genzel}, {Gillessen}, {Gordo}, {Greenbaum}, {Habibi},
  {Haubois}, {Hau{\ss}mann}, {Henning}, {Hippler}, {Horrobin}, {Hubert},
  {Jimenez Rosales}, {Jocou}, {Kendrew}, {Kervella}, {Kolb}, {Lagrange},
  {Lapeyr{\`e}re}, {Le Bouquin}, {L{\'e}na}, {Lippa}, {Lenzen}, {Maire},
  {Molli{\`e}re}, {Ott}, {Paumard}, {Perraut}, {Perrin}, {Pueyo}, {Rabien},
  {Ram{\'\i}rez}, {Rau}, {Rodr{\'\i}guez-Coira}, {Rousset}, {Sanchez-Bermudez},
  {Scheithauer}, {Schuhler}, {Straub}, {Straubmeier}, {Sturm}, {Tacconi},
  {Vincent}, {van Dishoeck}, {von Fellenberg}, {Wank}, {Waisberg}, {Widmann},
  {Wieprecht}, {Wiest}, {Wiezorrek}, {Woillez}, {Yazici}, {Ziegler}, \&
  {Zins}}]{GRAVITYCollaboration2019}
{GRAVITY Collaboration}, {Lacour}, S., {Nowak}, M., {et~al.} 2019,
  \href{http://dx.doi.org/10.1051/0004-6361/201935253}{\JournalTitle{\aap},
  623, L11}

\bibitem[{{Gravity Collaboration} {et~al.}(2020){Gravity Collaboration},
  {Nowak}, {Lacour}, {Molli{\`e}re}, {Wang}, {Charnay}, {van Dishoeck},
  {Abuter}, {Amorim}, {Berger}, {Beust}, {Bonnefoy}, {Bonnet}, {Brandner},
  {Buron}, {Cantalloube}, {Collin}, {Chapron}, {Cl{\'e}net}, {Coud{\'e} Du
  Foresto}, {de Zeeuw}, {Dembet}, {Dexter}, {Duvert}, {Eckart}, {Eisenhauer},
  {F{\"o}rster Schreiber}, {F{\'e}dou}, {Garcia Lopez}, {Gao}, {Gendron},
  {Genzel}, {Gillessen}, {Hau{\ss}mann}, {Henning}, {Hippler}, {Hubert},
  {Jocou}, {Kervella}, {Lagrange}, {Lapeyr{\`e}re}, {Le Bouquin}, {L{\'e}na},
  {Maire}, {Ott}, {Paumard}, {Paladini}, {Perraut}, {Perrin}, {Pueyo}, {Pfuhl},
  {Rabien}, {Rau}, {Rodr{\'\i}guez-Coira}, {Rousset}, {Scheithauer},
  {Shangguan}, {Straub}, {Straubmeier}, {Sturm}, {Tacconi}, {Vincent},
  {Widmann}, {Wieprecht}, {Wiezorrek}, {Woillez}, {Yazici}, \&
  {Ziegler}}]{GravityCollaboration2020}
{Gravity Collaboration}, {Nowak}, M., {Lacour}, S., {et~al.} 2020,
  \href{http://dx.doi.org/10.1051/0004-6361/201936898}{\JournalTitle{\aap},
  633, A110}

\bibitem[{{Green}(1985)}]{Green1985}
{Green}, R.~M. 1985, {Spherical Astronomy} (Cambridge University Press)

\bibitem[{{Greenbaum} {et~al.}(2018){Greenbaum}, {Pueyo}, {Ruffio}, {Wang}, {De
  Rosa}, {Aguilar}, {Rameau}, {Barman}, {Marois}, {Marley}, {Konopacky},
  {Rajan}, {Macintosh}, {Ansdell}, {Arriaga}, {Bailey}, {Bulger}, {Burrows},
  {Chilcote}, {Cotten}, {Doyon}, {Duch{\^e}ne}, {Fitzgerald}, {Follette},
  {Gerard}, {Goodsell}, {Graham}, {Hibon}, {Hung}, {Ingraham}, {Kalas},
  {Larkin}, {Maire}, {Marchis}, {Metchev}, {Millar-Blanchaer}, {Nielsen},
  {Norton}, {Oppenheimer}, {Palmer}, {Patience}, {Perrin}, {Poyneer},
  {Rantakyr{\"o}}, {Savransky}, {Schneider}, {Sivaramakrishnan}, {Song},
  {Soummer}, {Thomas}, {Wallace}, {Ward-Duong}, {Wiktorowicz}, \&
  {Wolff}}]{Greenbaum2018}
{Greenbaum}, A.~Z., {Pueyo}, L., {Ruffio}, J.-B., {et~al.} 2018,
  \href{http://dx.doi.org/10.3847/1538-3881/aabcb8}{\JournalTitle{\aj}, 155,
  226}

\bibitem[{{Greenbaum} {et~al.}(2023){Greenbaum}, {Llop-Sayson}, {Lew},
  {Bryden}, {Roellig}, {Ygouf}, {Fulton}, {Hey}, {Huber}, {Mukherjee}, {Meyer},
  {Leisenring}, {Rieke}, {Boyer}, {Green}, {Kelly}, {Misselt}, {Serabyn},
  {Stansberry}, {Chu}, {De Furio}, {Johnstone}, {Schlieder}, \&
  {Beichman}}]{Greenbaum2023}
{Greenbaum}, A.~Z., {Llop-Sayson}, J., {Lew}, B. W.~P., {et~al.} 2023,
  \href{http://dx.doi.org/10.3847/1538-4357/acb68b}{\JournalTitle{\apj}, 945,
  126}

\bibitem[{{Hinkley} {et~al.}(2015){Hinkley}, {Kraus}, {Ireland}, {Cheetham},
  {Carpenter}, {Tuthill}, {Lacour}, {Evans}, \& {Haubois}}]{Hinkley2015}
{Hinkley}, S., {Kraus}, A.~L., {Ireland}, M.~J., {et~al.} 2015,
  \href{http://dx.doi.org/10.1088/2041-8205/806/1/L9}{\JournalTitle{\apjl},
  806, L9}

\bibitem[{{Hinkley} {et~al.}(2023){Hinkley}, {Lacour}, {Marleau}, {Lagrange},
  {Wang}, {Kammerer}, {Cumming}, {Nowak}, {Rodet}, {Stolker}, {Balmer}, {Ray},
  {Bonnefoy}, {Molli{\`e}re}, {Lazzoni}, {Kennedy}, {Mordasini}, {Abuter},
  {Aigrain}, {Amorim}, {Asensio-Torres}, {Babusiaux}, {Benisty}, {Berger},
  {Beust}, {Blunt}, {Boccaletti}, {Bohn}, {Bonnet}, {Bourdarot}, {Brandner},
  {Cantalloube}, {Caselli}, {Charnay}, {Chauvin}, {Chomez}, {Choquet},
  {Christiaens}, {Cl{\'e}net}, {Coud{\'e} du Foresto}, {Cridland}, {Delorme},
  {Dembet}, {Drescher}, {Duvert}, {Eckart}, {Eisenhauer}, {Feuchtgruber},
  {Galland}, {Garcia}, {Garcia Lopez}, {Gardner}, {Gendron}, {Genzel},
  {Gillessen}, {Girard}, {Grandjean}, {Haubois}, {Hei{\ss}el}, {Henning},
  {Hippler}, {Horrobin}, {Houll{\'e}}, {Hubert}, {Jocou}, {Keppler},
  {Kervella}, {Kreidberg}, {Lapeyr{\`e}re}, {Le Bouquin}, {L{\'e}na}, {Lutz},
  {Maire}, {Mang}, {M{\'e}rand}, {Meunier}, {Monnier}, {Mouillet}, {Nasedkin},
  {Ott}, {Otten}, {Paladini}, {Paumard}, {Perraut}, {Perrin}, {Philipot},
  {Pfuhl}, {Pourr{\'e}}, {Pueyo}, {Rameau}, {Rickman}, {Rubini}, {Rustamkulov},
  {Samland}, {Shangguan}, {Shimizu}, {Sing}, {Straubmeier}, {Sturm}, {Tacconi},
  {van Dishoeck}, {Vigan}, {Vincent}, {Ward-Duong}, {Widmann}, {Wieprecht},
  {Wiezorrek}, {Woillez}, {Yazici}, {Young}, \& {Zicher}}]{Hinkley2023}
{Hinkley}, S., {Lacour}, S., {Marleau}, G.~D., {et~al.} 2023,
  \href{http://dx.doi.org/10.1051/0004-6361/202244727}{\JournalTitle{\aap},
  671, L5}

\bibitem[{{H{\o}g} {et~al.}(2000){H{\o}g}, {Fabricius}, {Makarov}, {Urban},
  {Corbin}, {Wycoff}, {Bastian}, {Schwekendiek}, \& {Wicenec}}]{Hog2000}
{H{\o}g}, E., {Fabricius}, C., {Makarov}, V.~V., {et~al.} 2000,
  \JournalTitle{\aap}, 355, L27

\bibitem[{{Houk}(1982)}]{Houk1982}
{Houk}, N. 1982, {Michigan Catalogue of Two-dimensional Spectral Types for the
  HD stars. Volume\_3. Declinations -40\_{\textflorin}0 to
  -26\_{\textflorin}0.}

\bibitem[{{Iyer} {et~al.}(2023){Iyer}, {Line}, {Muirhead}, {Fortney}, \&
  {Gharib-Nezhad}}]{Iyer2023}
{Iyer}, A.~R., {Line}, M.~R., {Muirhead}, P.~S., {Fortney}, J.~J., \&
  {Gharib-Nezhad}, E. 2023,
  \href{http://dx.doi.org/10.3847/1538-4357/acabc2}{\JournalTitle{\apj}, 944,
  41}

\bibitem[{Iyer {et~al.}(2022)Iyer, Line, Muirhead, Fortney, \&
  Gharib-Nezhad}]{iyer2022_data}
Iyer, R.~A., Line, R.~M., Muirhead, S.~P., Fortney, J.~J., \& Gharib-Nezhad, E.
  2022, {The SPHINX M-dwarf Spectral Grid. I. Benchmarking New Model
  Atmospheres to Derive Fundamental M-Dwarf Properties}

\bibitem[{{Janson} {et~al.}(2012){Janson}, {Jayawardhana}, {Girard},
  {Lafreni{\`e}re}, {Bonavita}, {Gizis}, \& {Brandeker}}]{Janson2012}
{Janson}, M., {Jayawardhana}, R., {Girard}, J.~H., {et~al.} 2012,
  \href{http://dx.doi.org/10.1088/2041-8205/758/1/L2}{\JournalTitle{\apjl},
  758, L2}

\bibitem[{{Kammerer} {et~al.}(2021){Kammerer}, {Lacour}, {Stolker},
  {Molli{\`e}re}, {Sing}, {Nasedkin}, {Kervella}, {Wang}, {Ward-Duong},
  {Nowak}, {Abuter}, {Amorim}, {Asensio-Torres}, {Baub{\"o}ck}, {Benisty},
  {Berger}, {Beust}, {Blunt}, {Boccaletti}, {Bohn}, {Bolzer}, {Bonnefoy},
  {Bonnet}, {Brandner}, {Cantalloube}, {Caselli}, {Charnay}, {Chauvin},
  {Choquet}, {Christiaens}, {Cl{\'e}net}, {Coud{\'e} du Foresto}, {Cridland},
  {Dembet}, {Dexter}, {de Zeeuw}, {Drescher}, {Duvert}, {Eckart}, {Eisenhauer},
  {Gao}, {Garcia}, {Garcia Lopez}, {Gendron}, {Genzel}, {Gillessen}, {Girard},
  {Haubois}, {Hei{\ss}el}, {Henning}, {Hinkley}, {Hippler}, {Horrobin},
  {Houll{\'e}}, {Hubert}, {Jocou}, {Keppler}, {Kreidberg}, {Lagrange},
  {Lapeyr{\`e}re}, {Le Bouquin}, {L{\'e}na}, {Lutz}, {Maire}, {M{\'e}rand},
  {Monnier}, {Mouillet}, {M{\"u}ller}, {Ott}, {Otten}, {Paladini}, {Paumard},
  {Perraut}, {Perrin}, {Pfuhl}, {Pueyo}, {Rameau}, {Rodet}, {Rousset},
  {Rustamkulov}, {Shangguan}, {Shimizu}, {Stadler}, {Straub}, {Straubmeier},
  {Sturm}, {Tacconi}, {van Dishoeck}, {Vigan}, {Vincent}, {von Fellenberg},
  {Widmann}, {Wieprecht}, {Wiezorrek}, {Woillez}, \& {Yazici}}]{Kammerer2021}
{Kammerer}, J., {Lacour}, S., {Stolker}, T., {et~al.} 2021,
  \href{http://dx.doi.org/10.1051/0004-6361/202140749}{\JournalTitle{\aap},
  652, A57}

\bibitem[{{Kervella} {et~al.}(2019){Kervella}, {Arenou}, {Mignard}, \&
  {Th{\'e}venin}}]{Kervella2019}
{Kervella}, P., {Arenou}, F., {Mignard}, F., \& {Th{\'e}venin}, F. 2019,
  \href{http://dx.doi.org/10.1051/0004-6361/201834371}{\JournalTitle{\aap},
  623, A72}

\bibitem[{{Kervella} {et~al.}(2022){Kervella}, {Arenou}, \&
  {Th{\'e}venin}}]{Kervella2022}
{Kervella}, P., {Arenou}, F., \& {Th{\'e}venin}, F. 2022,
  \href{http://dx.doi.org/10.1051/0004-6361/202142146}{\JournalTitle{\aap},
  657, A7}

\bibitem[{{Kouwenhoven} {et~al.}(2007){Kouwenhoven}, {Brown}, \&
  {Kaper}}]{Kouwenhoven2007}
{Kouwenhoven}, M.~B.~N., {Brown}, A.~G.~A., \& {Kaper}, L. 2007,
  \href{http://dx.doi.org/10.1051/0004-6361:20054396}{\JournalTitle{\aap}, 464,
  581}

\bibitem[{{Kratter} {et~al.}(2010){Kratter}, {Murray-Clay}, \&
  {Youdin}}]{Kratter2010}
{Kratter}, K.~M., {Murray-Clay}, R.~A., \& {Youdin}, A.~N. 2010,
  \href{http://dx.doi.org/10.1088/0004-637X/710/2/1375}{\JournalTitle{\apj},
  710, 1375}

\bibitem[{{Lacour} {et~al.}(2019){Lacour}, {Dembet}, {Abuter}, {F{\'e}dou},
  {Perrin}, {Choquet}, {Pfuhl}, {Eisenhauer}, {Woillez}, {Cassaing},
  {Wieprecht}, {Ott}, {Wiezorrek}, {Tristram}, {Wolff}, {Ram{\'\i}rez},
  {Haubois}, {Perraut}, {Straubmeier}, {Brandner}, \& {Amorim}}]{Lacour2019}
{Lacour}, S., {Dembet}, R., {Abuter}, R., {et~al.} 2019,
  \href{http://dx.doi.org/10.1051/0004-6361/201834981}{\JournalTitle{\aap},
  624, A99}

\bibitem[{{Lacour} {et~al.}(2021){Lacour}, {Wang}, {Rodet}, {Nowak},
  {Shangguan}, {Beust}, {Lagrange}, {Abuter}, {Amorim}, {Asensio-Torres},
  {Benisty}, {Berger}, {Blunt}, {Boccaletti}, {Bohn}, {Bolzer}, {Bonnefoy},
  {Bonnet}, {Bourdarot}, {Brandner}, {Cantalloube}, {Caselli}, {Charnay},
  {Chauvin}, {Choquet}, {Christiaens}, {Cl{\'e}net}, {Coud{\'e} Du Foresto},
  {Cridland}, {Dembet}, {Dexter}, {de Zeeuw}, {Drescher}, {Duvert}, {Eckart},
  {Eisenhauer}, {Gao}, {Garcia}, {Garcia Lopez}, {Gendron}, {Genzel},
  {Gillessen}, {Girard}, {Haubois}, {Hei{\ss}el}, {Henning}, {Hinkley},
  {Hippler}, {Horrobin}, {Houll{\'e}}, {Hubert}, {Jocou}, {Kammerer},
  {Keppler}, {Kervella}, {Kreidberg}, {Lapeyr{\`e}re}, {Le Bouquin},
  {L{\'e}na}, {Lutz}, {Maire}, {M{\'e}rand}, {Molli{\`e}re}, {Monnier},
  {Mouillet}, {Nasedkin}, {Ott}, {Otten}, {Paladini}, {Paumard}, {Perraut},
  {Perrin}, {Pfuhl}, {Rickman}, {Pueyo}, {Rameau}, {Rousset}, {Rustamkulov},
  {Samland}, {Shimizu}, {Sing}, {Stadler}, {Stolker}, {Straub}, {Straubmeier},
  {Sturm}, {Tacconi}, {van Dishoeck}, {Vigan}, {Vincent}, {von Fellenberg},
  {Ward-Duong}, {Widmann}, {Wieprecht}, {Wiezorrek}, {Woillez}, {Yazici},
  {Young}, \& {Gravity Collaboration}}]{Lacour2021}
{Lacour}, S., {Wang}, J.~J., {Rodet}, L., {et~al.} 2021,
  \href{http://dx.doi.org/10.1051/0004-6361/202141889}{\JournalTitle{\aap},
  654, L2}

\bibitem[{{Lagrange} {et~al.}(2020){Lagrange}, {Rubini}, {Nowak}, {Lacour},
  {Grandjean}, {Boccaletti}, {Langlois}, {Delorme}, {Gratton}, {Wang},
  {Flasseur}, {Galicher}, {Kral}, {Meunier}, {Beust}, {Babusiaux}, {Le
  Coroller}, {Thebault}, {Kervella}, {Zurlo}, {Maire}, {Wahhaj}, {Amorim},
  {Asensio-Torres}, {Benisty}, {Berger}, {Bonnefoy}, {Brandner}, {Cantalloube},
  {Charnay}, {Chauvin}, {Choquet}, {Cl{\'e}net}, {Christiaens}, {Coud{\'e} Du
  Foresto}, {de Zeeuw}, {Desidera}, {Duvert}, {Eckart}, {Eisenhauer},
  {Galland}, {Gao}, {Garcia}, {Garcia Lopez}, {Gendron}, {Genzel}, {Gillessen},
  {Girard}, {Hagelberg}, {Haubois}, {Henning}, {Heissel}, {Hippler},
  {Horrobin}, {Janson}, {Kammerer}, {Kenworthy}, {Keppler}, {Kreidberg},
  {Lapeyr{\`e}re}, {Le Bouquin}, {L{\'e}na}, {M{\'e}rand}, {Messina},
  {Molli{\`e}re}, {Monnier}, {Ott}, {Otten}, {Paumard}, {Paladini}, {Perraut},
  {Perrin}, {Pueyo}, {Pfuhl}, {Rodet}, {Rodriguez-Coira}, {Rousset}, {Samland},
  {Shangguan}, {Schmidt}, {Straub}, {Straubmeier}, {Stolker}, {Vigan},
  {Vincent}, {Widmann}, {Woillez}, \& {GRAVITY Collaboration}}]{Lagrange2020}
{Lagrange}, A.~M., {Rubini}, P., {Nowak}, M., {et~al.} 2020,
  \href{http://dx.doi.org/10.1051/0004-6361/202038823}{\JournalTitle{\aap},
  642, A18}

\bibitem[{{Lapeyrere} {et~al.}(2014){Lapeyrere}, {Kervella}, {Lacour},
  {Azouaoui}, {Garcia-Dabo}, {Perrin}, {Eisenhauer}, {Perraut}, {Straubmeier},
  {Amorim}, \& {Brandner}}]{Lapeyrere2014}
{Lapeyrere}, V., {Kervella}, P., {Lacour}, S., {et~al.} 2014,
  \href{http://dx.doi.org/10.1117/12.2056850}{in Society of Photo-Optical
  Instrumentation Engineers (SPIE) Conference Series, Vol. 9146, Optical and
  Infrared Interferometry IV, ed. J.~K. {Rajagopal}, M.~J. {Creech-Eakman}, \&
  F.~{Malbet}}, 91462D

\bibitem[{{Maire} {et~al.}(2016){Maire}, {Bonnefoy}, {Ginski}, {Vigan},
  {Messina}, {Mesa}, {Galicher}, {Gratton}, {Desidera}, {Kopytova}, {Millward},
  {Thalmann}, {Claudi}, {Ehrenreich}, {Zurlo}, {Chauvin}, {Antichi},
  {Baruffolo}, {Bazzon}, {Beuzit}, {Blanchard}, {Boccaletti}, {de Boer},
  {Carle}, {Cascone}, {Costille}, {De Caprio}, {Delboulb{\'e}}, {Dohlen},
  {Dominik}, {Feldt}, {Fusco}, {Girard}, {Giro}, {Gisler}, {Gluck}, {Gry},
  {Henning}, {Hubin}, {Hugot}, {Jaquet}, {Kasper}, {Lagrange}, {Langlois}, {Le
  Mignant}, {Llored}, {Madec}, {Martinez}, {Mawet}, {Milli},
  {M{\"o}ller-Nilsson}, {Mouillet}, {Moulin}, {Moutou}, {Orign{\'e}}, {Pavlov},
  {Petit}, {Pragt}, {Puget}, {Ramos}, {Rochat}, {Roelfsema}, {Salasnich},
  {Sauvage}, {Schmid}, {Turatto}, {Udry}, {Vakili}, {Wahhaj}, {Weber}, \&
  {Wildi}}]{Maire2016}
{Maire}, A.~L., {Bonnefoy}, M., {Ginski}, C., {et~al.} 2016,
  \href{http://dx.doi.org/10.1051/0004-6361/201526594}{\JournalTitle{\aap},
  587, A56}

\bibitem[{{Males} {et~al.}(2022){Males}, {Close}, {Haffert}, {Long}, {Hedglen},
  {Pearce}, {Weinberger}, {Guyon}, {Knight}, {McLeod}, {Kautz}, {Van Gorkom},
  {Lumbres}, {Schatz}, {Rodack}, {Gasho}, {Kueny}, \& {Foster}}]{Males2022}
{Males}, J.~R., {Close}, L.~M., {Haffert}, S., {et~al.} 2022,
  \href{http://dx.doi.org/10.1117/12.2630584}{in Society of Photo-Optical
  Instrumentation Engineers (SPIE) Conference Series, Vol. 12185, Adaptive
  Optics Systems VIII, ed. L.~{Schreiber}, D.~{Schmidt}, \& E.~{Vernet}},
  1218509

\bibitem[{{Milli} {et~al.}(2017){Milli}, {Hibon}, {Christiaens}, {Choquet},
  {Bonnefoy}, {Kennedy}, {Wyatt}, {Absil}, {G{\'o}mez Gonz{\'a}lez}, {del
  Burgo}, {Matr{\`a}}, {Augereau}, {Boccaletti}, {Delacroix}, {Ertel}, {Dent},
  {Forsberg}, {Fusco}, {Girard}, {Habraken}, {Huby}, {Karlsson}, {Lagrange},
  {Mawet}, {Mouillet}, {Perrin}, {Pinte}, {Pueyo}, {Reyes}, {Soummer},
  {Surdej}, {Tarricq}, \& {Wahhaj}}]{Milli2017}
{Milli}, J., {Hibon}, P., {Christiaens}, V., {et~al.} 2017,
  \href{http://dx.doi.org/10.1051/0004-6361/201629908}{\JournalTitle{\aap},
  597, L2}

\bibitem[{{Molli{\`e}re} \& {Mordasini}(2012)}]{Molliere2012}
{Molli{\`e}re}, P., \& {Mordasini}, C. 2012,
  \href{http://dx.doi.org/10.1051/0004-6361/201219844}{\JournalTitle{\aap},
  547, A105}

\bibitem[{{Molli{\`e}re} {et~al.}(2020){Molli{\`e}re}, {Stolker}, {Lacour},
  {Otten}, {Shangguan}, {Charnay}, {Molyarova}, {Nowak}, {Henning}, {Marleau},
  {Semenov}, {van Dishoeck}, {Eisenhauer}, {Garcia}, {Garcia Lopez}, {Girard},
  {Greenbaum}, {Hinkley}, {Kervella}, {Kreidberg}, {Maire}, {Nasedkin},
  {Pueyo}, {Snellen}, {Vigan}, {Wang}, {de Zeeuw}, \& {Zurlo}}]{Molliere2020}
{Molli{\`e}re}, P., {Stolker}, T., {Lacour}, S., {et~al.} 2020,
  \href{http://dx.doi.org/10.1051/0004-6361/202038325}{\JournalTitle{\aap},
  640, A131}

\bibitem[{{Mordasini} {et~al.}(2009){Mordasini}, {Alibert}, \&
  {Benz}}]{Mordasini2009}
{Mordasini}, C., {Alibert}, Y., \& {Benz}, W. 2009,
  \href{http://dx.doi.org/10.1051/0004-6361/200810301}{\JournalTitle{\aap},
  501, 1139}

\bibitem[{{Mordasini} {et~al.}(2012){Mordasini}, {Alibert}, {Klahr}, \&
  {Henning}}]{Mordasini2012}
{Mordasini}, C., {Alibert}, Y., {Klahr}, H., \& {Henning}, T. 2012,
  \href{http://dx.doi.org/10.1051/0004-6361/201118457}{\JournalTitle{\aap},
  547, A111}

\bibitem[{{Mugrauer} {et~al.}(2010){Mugrauer}, {Vogt}, {Neuh{\"a}user}, \&
  {Schmidt}}]{Mugrauer2010}
{Mugrauer}, M., {Vogt}, N., {Neuh{\"a}user}, R., \& {Schmidt}, T.~O.~B. 2010,
  \href{http://dx.doi.org/10.1051/0004-6361/201015523}{\JournalTitle{\aap},
  523, L1}

\bibitem[{{Nagpal} {et~al.}(2023){Nagpal}, {Blunt}, {Bowler}, {Dupuy},
  {Nielsen}, \& {Wang}}]{Nagpal2023}
{Nagpal}, V., {Blunt}, S., {Bowler}, B.~P., {et~al.} 2023,
  \href{http://dx.doi.org/10.3847/1538-3881/ac9fd2}{\JournalTitle{\aj}, 165,
  32}

\bibitem[{{Nielsen} {et~al.}(2019){Nielsen}, {De Rosa}, {Macintosh}, {Wang},
  {Ruffio}, {Chiang}, {Marley}, {Saumon}, {Savransky}, {Ammons}, {Bailey},
  {Barman}, {Blain}, {Bulger}, {Burrows}, {Chilcote}, {Cotten}, {Czekala},
  {Doyon}, {Duch{\^e}ne}, {Esposito}, {Fabrycky}, {Fitzgerald}, {Follette},
  {Fortney}, {Gerard}, {Goodsell}, {Graham}, {Greenbaum}, {Hibon}, {Hinkley},
  {Hirsch}, {Hom}, {Hung}, {Dawson}, {Ingraham}, {Kalas}, {Konopacky},
  {Larkin}, {Lee}, {Lin}, {Maire}, {Marchis}, {Marois}, {Metchev},
  {Millar-Blanchaer}, {Morzinski}, {Oppenheimer}, {Palmer}, {Patience},
  {Perrin}, {Poyneer}, {Pueyo}, {Rafikov}, {Rajan}, {Rameau}, {Rantakyr{\"o}},
  {Ren}, {Schneider}, {Sivaramakrishnan}, {Song}, {Soummer}, {Tallis},
  {Thomas}, {Ward-Duong}, \& {Wolff}}]{Nielsen2019}
{Nielsen}, E.~L., {De Rosa}, R.~J., {Macintosh}, B., {et~al.} 2019,
  \href{http://dx.doi.org/10.3847/1538-3881/ab16e9}{\JournalTitle{\aj}, 158,
  13}

\bibitem[{{Nowak} {et~al.}(2020){Nowak}, {Lacour}, {Lagrange}, {Rubini},
  {Wang}, {Stolker}, {Abuter}, {Amorim}, {Asensio-Torres}, {Baub{\"o}ck},
  {Benisty}, {Berger}, {Beust}, {Blunt}, {Boccaletti}, {Bonnefoy}, {Bonnet},
  {Brandner}, {Cantalloube}, {Charnay}, {Choquet}, {Christiaens}, {Cl{\'e}net},
  {Coud{\'e} Du Foresto}, {Cridland}, {de Zeeuw}, {Dembet}, {Dexter},
  {Drescher}, {Duvert}, {Eckart}, {Eisenhauer}, {Gao}, {Garcia}, {Garcia
  Lopez}, {Gardner}, {Gendron}, {Genzel}, {Gillessen}, {Girard}, {Grandjean},
  {Haubois}, {Hei{\ss}el}, {Henning}, {Hinkley}, {Hippler}, {Horrobin},
  {Houll{\'e}}, {Hubert}, {Jim{\'e}nez-Rosales}, {Jocou}, {Kammerer},
  {Kervella}, {Keppler}, {Kreidberg}, {Kulikauskas}, {Lapeyr{\`e}re}, {Le
  Bouquin}, {L{\'e}na}, {M{\'e}rand}, {Maire}, {Molli{\`e}re}, {Monnier},
  {Mouillet}, {M{\"u}ller}, {Nasedkin}, {Ott}, {Otten}, {Paumard}, {Paladini},
  {Perraut}, {Perrin}, {Pueyo}, {Pfuhl}, {Rameau}, {Rodet},
  {Rodr{\'\i}guez-Coira}, {Rousset}, {Scheithauer}, {Shangguan}, {Stadler},
  {Straub}, {Straubmeier}, {Sturm}, {Tacconi}, {van Dishoeck}, {Vigan},
  {Vincent}, {von Fellenberg}, {Ward-Duong}, {Widmann}, {Wieprecht},
  {Wiezorrek}, {Woillez}, \& {GRAVITY Collaboration}}]{Nowak2020}
{Nowak}, M., {Lacour}, S., {Lagrange}, A.~M., {et~al.} 2020,
  \href{http://dx.doi.org/10.1051/0004-6361/202039039}{\JournalTitle{\aap},
  642, L2}

\bibitem[{{Padoan} \& {Nordlund}(2004)}]{Padoan2004}
{Padoan}, P., \& {Nordlund}, {\r{A}}. 2004,
  \href{http://dx.doi.org/10.1086/345413}{\JournalTitle{\apj}, 617, 559}

\bibitem[{{Parker} \& {Daffern-Powell}(2022)}]{Parker2022}
{Parker}, R.~J., \& {Daffern-Powell}, E.~C. 2022,
  \href{http://dx.doi.org/10.1093/mnrasl/slac086}{\JournalTitle{\mnras}, 516,
  L91}

\bibitem[{{Pecaut} \& {Mamajek}(2016)}]{Pecaut2016}
{Pecaut}, M.~J., \& {Mamajek}, E.~E. 2016,
  \href{http://dx.doi.org/10.1093/mnras/stw1300}{\JournalTitle{\mnras}, 461,
  794}

\bibitem[{{Petrus} {et~al.}(2021){Petrus}, {Bonnefoy}, {Chauvin}, {Charnay},
  {Marleau}, {Gratton}, {Lagrange}, {Rameau}, {Mordasini}, {Nowak}, {Delorme},
  {Boccaletti}, {Carlotti}, {Houll{\'e}}, {Vigan}, {Allard}, {Desidera},
  {D'Orazi}, {Hoeijmakers}, {Wyttenbach}, \& {Lavie}}]{Petrus2021}
{Petrus}, S., {Bonnefoy}, M., {Chauvin}, G., {et~al.} 2021,
  \href{http://dx.doi.org/10.1051/0004-6361/202038914}{\JournalTitle{\aap},
  648, A59}

\bibitem[{{Phillips} {et~al.}(2020){Phillips}, {Tremblin}, {Baraffe},
  {Chabrier}, {Allard}, {Spiegelman}, {Goyal}, {Drummond}, \&
  {H{\'e}brard}}]{Phillips2020}
{Phillips}, M.~W., {Tremblin}, P., {Baraffe}, I., {et~al.} 2020,
  \href{http://dx.doi.org/10.1051/0004-6361/201937381}{\JournalTitle{\aap},
  637, A38}

\bibitem[{{Pollack} {et~al.}(1996){Pollack}, {Hubickyj}, {Bodenheimer},
  {Lissauer}, {Podolak}, \& {Greenzweig}}]{Pollack1996}
{Pollack}, J.~B., {Hubickyj}, O., {Bodenheimer}, P., {et~al.} 1996,
  \href{http://dx.doi.org/10.1006/icar.1996.0190}{\JournalTitle{\icarus}, 124,
  62}

\bibitem[{{Pueyo}(2016)}]{Pueyo2016}
{Pueyo}, L. 2016,
  \href{http://dx.doi.org/10.3847/0004-637X/824/2/117}{\JournalTitle{\apj},
  824, 117}

\bibitem[{{Rickman} {et~al.}(2020){Rickman}, {S{\'e}gransan}, {Hagelberg},
  {Beuzit}, {Cheetham}, {Delisle}, {Forveille}, \& {Udry}}]{Rickman2020}
{Rickman}, E.~L., {S{\'e}gransan}, D., {Hagelberg}, J., {et~al.} 2020,
  \href{http://dx.doi.org/10.1051/0004-6361/202037524}{\JournalTitle{\aap},
  635, A203}

\bibitem[{{Rickman} {et~al.}(2022){Rickman}, {Matthews}, {Ceva},
  {S{\'e}gransan}, {Brandt}, {Zhang}, {Brandt}, {Forveille}, {Hagelberg}, \&
  {Udry}}]{Rickman2022}
{Rickman}, E.~L., {Matthews}, E., {Ceva}, W., {et~al.} 2022,
  \href{http://dx.doi.org/10.1051/0004-6361/202244633}{\JournalTitle{\aap},
  668, A140}

\bibitem[{{Rodet} {et~al.}(2018){Rodet}, {Bonnefoy}, {Durkan}, {Beust},
  {Lagrange}, {Schlieder}, {Janson}, {Grandjean}, {Chauvin}, {Messina},
  {Maire}, {Brandner}, {Girard}, {Delorme}, {Biller}, {Bergfors}, {Lacour},
  {Feldt}, {Henning}, {Boccaletti}, {Le Bouquin}, {Berger}, {Monin}, {Udry},
  {Peretti}, {Segransan}, {Allard}, {Homeier}, {Vigan}, {Langlois},
  {Hagelberg}, {Menard}, {Bazzon}, {Beuzit}, {Delboulb{\'e}}, {Desidera},
  {Gratton}, {Lannier}, {Ligi}, {Maurel}, {Mesa}, {Meyer}, {Pavlov}, {Ramos},
  {Rigal}, {Roelfsema}, {Salter}, {Samland}, {Schmidt}, {Stadler}, \&
  {Weber}}]{Rodet2018}
{Rodet}, L., {Bonnefoy}, M., {Durkan}, S., {et~al.} 2018,
  \href{http://dx.doi.org/10.1051/0004-6361/201832924}{\JournalTitle{\aap},
  618, A23}

\bibitem[{{Sanghi} {et~al.}(2023){Sanghi}, {Liu}, {Best}, {Dupuy}, {Siverd},
  {Zhang}, {Hurt}, {Magnier}, {Aller}, \& {Deacon}}]{Sanghi2023}
{Sanghi}, A., {Liu}, M.~C., {Best}, W.~M., {et~al.} 2023,
  \href{http://dx.doi.org/10.48550/arXiv.2309.03082}{\JournalTitle{arXiv
  e-prints}, arXiv:2309.03082}

\bibitem[{{Saumon} {et~al.}(1996){Saumon}, {Hubbard}, {Burrows}, {Guillot},
  {Lunine}, \& {Chabrier}}]{Saumon1996}
{Saumon}, D., {Hubbard}, W.~B., {Burrows}, A., {et~al.} 1996,
  \href{http://dx.doi.org/10.1086/177027}{\JournalTitle{\apj}, 460, 993}

\bibitem[{{Schlaufman}(2018)}]{Schlaufman2018}
{Schlaufman}, K.~C. 2018,
  \href{http://dx.doi.org/10.3847/1538-4357/aa961c}{\JournalTitle{\apj}, 853,
  37}

\bibitem[{{Schmidt} {et~al.}(2014){Schmidt}, {Mugrauer}, {Neuh{\"a}user},
  {Vogt}, {Witte}, {Hauschildt}, {Helling}, \& {Seifahrt}}]{Schmidt2014}
{Schmidt}, T.~O.~B., {Mugrauer}, M., {Neuh{\"a}user}, R., {et~al.} 2014,
  \href{http://dx.doi.org/10.1051/0004-6361/201321625}{\JournalTitle{\aap},
  566, A85}

\bibitem[{{Soummer} {et~al.}(2012){Soummer}, {Pueyo}, \&
  {Larkin}}]{Soummer2012}
{Soummer}, R., {Pueyo}, L., \& {Larkin}, J. 2012,
  \href{http://dx.doi.org/10.1088/2041-8205/755/2/L28}{\JournalTitle{\apjl},
  755, L28}

\bibitem[{{Squicciarini} {et~al.}(2022){Squicciarini}, {Gratton}, {Janson},
  {Mamajek}, {Chauvin}, {Delorme}, {Langlois}, {Vigan}, {Ringqvist}, {Meeus},
  {Reffert}, {Kenworthy}, {Meyer}, {Bonnefoy}, {Bonavita}, {Mesa}, {Samland},
  {Desidera}, {D'Orazi}, {Engler}, {Alecian}, {Miglio}, {Henning}, {Quanz},
  {Mayer}, {Flasseur}, \& {Marleau}}]{Squicciarini2022}
{Squicciarini}, V., {Gratton}, R., {Janson}, M., {et~al.} 2022,
  \href{http://dx.doi.org/10.1051/0004-6361/202243675}{\JournalTitle{\aap},
  664, A9}

\bibitem[{{Stamatellos} {et~al.}(2007){Stamatellos}, {Hubber}, \&
  {Whitworth}}]{Stamatellos2007}
{Stamatellos}, D., {Hubber}, D.~A., \& {Whitworth}, A.~P. 2007,
  \href{http://dx.doi.org/10.1111/j.1745-3933.2007.00383.x}{\JournalTitle{\mnras},
  382, L30}

\bibitem[{{Stamatellos} \& {Whitworth}(2009)}]{Stamatellos2009}
{Stamatellos}, D., \& {Whitworth}, A.~P. 2009,
  \href{http://dx.doi.org/10.1111/j.1365-2966.2008.14069.x}{\JournalTitle{\mnras},
  392, 413}

\bibitem[{{Stolker} {et~al.}(2020){Stolker}, {Quanz}, {Todorov}, {K{\"u}hn},
  {Molli{\`e}re}, {Meyer}, {Currie}, {Daemgen}, \& {Lavie}}]{Stolker2020}
{Stolker}, T., {Quanz}, S.~P., {Todorov}, K.~O., {et~al.} 2020,
  \href{http://dx.doi.org/10.1051/0004-6361/201937159}{\JournalTitle{\aap},
  635, A182}

\bibitem[{{Vigan}(2020)}]{Vigan2020}
{Vigan}, A. 2020, {vlt-sphere: Automatic VLT/SPHERE data reduction and
  analysis}, Astrophysics Source Code Library, record ascl:2009.002

\bibitem[{{Vigan} {et~al.}(2010){Vigan}, {Moutou}, {Langlois}, {Allard},
  {Boccaletti}, {Carbillet}, {Mouillet}, \& {Smith}}]{Vigan2010}
{Vigan}, A., {Moutou}, C., {Langlois}, M., {et~al.} 2010,
  \href{http://dx.doi.org/10.1111/j.1365-2966.2010.16916.x}{\JournalTitle{\mnras},
  407, 71}

\bibitem[{{Vigan} {et~al.}(2023){Vigan}, {El Morsy}, {Lopez}, {Otten},
  {Garcia}, {Costes}, {Muslimov}, {Viret}, {Charles}, {Zins}, {Murray},
  {Costille}, {Paufique}, {Seemann}, {Houll{\'e}}, {Anwand-Heerwart},
  {Phillips}, {Abinanti}, {Balard}, {Baraffe}, {Benedetti}, {Blanchard},
  {Blanco}, {Beuzit}, {Choquet}, {Cristofari}, {Desidera}, {Dohlen}, {Dorn},
  {Ely}, {Fuenteseca}, {Garcia}, {Jaquet}, {Jaubert}, {Kasper}, {Le Merrer},
  {Maire}, {N'Diaye}, {Pallanca}, {Popovic}, {Pourcelot}, {Reiners}, {Rochat},
  {Sehim}, {Schmutzer}, {Smette}, {Tchoubaklian}, {Tomlinson}, \& {Valenzuela
  Soto}}]{Vigan2023}
{Vigan}, A., {El Morsy}, M., {Lopez}, M., {et~al.} 2023,
  \href{http://dx.doi.org/10.48550/arXiv.2309.12390}{\JournalTitle{arXiv
  e-prints}, arXiv:2309.12390}

\bibitem[{{Viswanath} {et~al.}(2023){Viswanath}, {Janson}, {Gratton},
  {Squicciarini}, {Rodet}, {Ringqvist}, {Mamajek}, {Reffert}, {Chauvin},
  {Delorme}, {Vigan}, {Bonnefoy}, {Engler}, {Desidera}, {Henning}, {Hagelberg},
  {Langlois}, \& {Meyer}}]{Viswanath2023}
{Viswanath}, G., {Janson}, M., {Gratton}, R., {et~al.} 2023,
  \href{http://dx.doi.org/10.1051/0004-6361/202346154}{\JournalTitle{\aap},
  675, A54}

\bibitem[{{Vousden} {et~al.}(2016){Vousden}, {Farr}, \& {Mandel}}]{Vousden2016}
{Vousden}, W.~D., {Farr}, W.~M., \& {Mandel}, I. 2016,
  \href{http://dx.doi.org/10.1093/mnras/stv2422}{\JournalTitle{\mnras}, 455,
  1919}

\bibitem[{{Vrijmoet} {et~al.}(2020){Vrijmoet}, {Henry}, {Jao}, \&
  {Dieterich}}]{Vrijmoet2020}
{Vrijmoet}, E.~H., {Henry}, T.~J., {Jao}, W.-C., \& {Dieterich}, S.~B. 2020,
  \href{http://dx.doi.org/10.3847/1538-3881/abb4e9}{\JournalTitle{\aj}, 160,
  215}

\bibitem[{{Vrijmoet} {et~al.}(2022){Vrijmoet}, {Tokovinin}, {Henry}, {Winters},
  {Horch}, \& {Jao}}]{Vrijmoet2022}
{Vrijmoet}, E.~H., {Tokovinin}, A., {Henry}, T.~J., {et~al.} 2022,
  \href{http://dx.doi.org/10.3847/1538-3881/ac52f6}{\JournalTitle{\aj}, 163,
  178}

\bibitem[{{{\v{Z}}erjal} {et~al.}(2023){{\v{Z}}erjal}, {Ireland}, {Crundall},
  {Krumholz}, \& {Rains}}]{Zerjal2023}
{{\v{Z}}erjal}, M., {Ireland}, M.~J., {Crundall}, T.~D., {Krumholz}, M.~R., \&
  {Rains}, A.~D. 2023,
  \href{http://dx.doi.org/10.1093/mnras/stac3693}{\JournalTitle{\mnras}, 519,
  3992}

\bibitem[{{Wagner} {et~al.}(2022){Wagner}, {Apai}, {Kasper}, {McClure}, \&
  {Robberto}}]{Wagner2022}
{Wagner}, K., {Apai}, D., {Kasper}, M., {McClure}, M., \& {Robberto}, M. 2022,
  \href{http://dx.doi.org/10.3847/1538-3881/ac409d}{\JournalTitle{\aj}, 163,
  80}

\bibitem[{{Wagner} {et~al.}(2020){Wagner}, {Apai}, {Kasper}, {McClure},
  {Robberto}, \& {Currie}}]{Wagner2020}
{Wagner}, K., {Apai}, D., {Kasper}, M., {et~al.} 2020,
  \href{http://dx.doi.org/10.3847/2041-8213/abb94e}{\JournalTitle{\apjl}, 902,
  L6}

\bibitem[{{Wang} {et~al.}(2015){Wang}, {Ruffio}, {De Rosa}, {Aguilar}, {Wolff},
  \& {Pueyo}}]{wang2015}
{Wang}, J.~J., {Ruffio}, J.-B., {De Rosa}, R.~J., {et~al.} 2015, {pyKLIP: PSF
  Subtraction for Exoplanets and Disks}, Astrophysics Source Code Library,
  record ascl:1506.001

\bibitem[{{Wang} {et~al.}(2016){Wang}, {Graham}, {Pueyo}, {Kalas},
  {Millar-Blanchaer}, {Ruffio}, {De Rosa}, {Ammons}, {Arriaga}, {Bailey},
  {Barman}, {Bulger}, {Burrows}, {Cardwell}, {Chen}, {Chilcote}, {Cotten},
  {Fitzgerald}, {Follette}, {Doyon}, {Duch{\^e}ne}, {Greenbaum}, {Hibon},
  {Hung}, {Ingraham}, {Konopacky}, {Larkin}, {Macintosh}, {Maire}, {Marchis},
  {Marley}, {Marois}, {Metchev}, {Nielsen}, {Oppenheimer}, {Palmer}, {Patel},
  {Patience}, {Perrin}, {Poyneer}, {Rajan}, {Rameau}, {Rantakyr{\"o}},
  {Savransky}, {Sivaramakrishnan}, {Song}, {Soummer}, {Thomas}, {Vasisht},
  {Vega}, {Wallace}, {Ward-Duong}, {Wiktorowicz}, \& {Wolff}}]{Wang2016}
{Wang}, J.~J., {Graham}, J.~R., {Pueyo}, L., {et~al.} 2016,
  \href{http://dx.doi.org/10.3847/0004-6256/152/4/97}{\JournalTitle{\aj}, 152,
  97}

\bibitem[{{Wang} {et~al.}(2020){Wang}, {Ginzburg}, {Ren}, {Wallack}, {Gao},
  {Mawet}, {Bond}, {Cetre}, {Wizinowich}, {De Rosa}, {Ruane}, {Liu}, {Absil},
  {Alvarez}, {Baranec}, {Choquet}, {Chun}, {Defr{\`e}re}, {Delorme},
  {Duch{\^e}ne}, {Forsberg}, {Ghez}, {Guyon}, {Hall}, {Huby}, {Jolivet},
  {Jensen-Clem}, {Jovanovic}, {Karlsson}, {Lilley}, {Matthews}, {M{\'e}nard},
  {Meshkat}, {Millar-Blanchaer}, {Ngo}, {Orban de Xivry}, {Pinte}, {Ragland},
  {Serabyn}, {Catal{\'a}n}, {Wang}, {Wetherell}, {Williams}, {Ygouf}, \&
  {Zuckerman}}]{Wang2020}
{Wang}, J.~J., {Ginzburg}, S., {Ren}, B., {et~al.} 2020,
  \href{http://dx.doi.org/10.3847/1538-3881/ab8aef}{\JournalTitle{\aj}, 159,
  263}

\bibitem[{{Wang} {et~al.}(2021){Wang}, {Vigan}, {Lacour}, {Nowak}, {Stolker},
  {De Rosa}, {Ginzburg}, {Gao}, {Abuter}, {Amorim}, {Asensio-Torres},
  {Baub{\"o}ck}, {Benisty}, {Berger}, {Beust}, {Beuzit}, {Blunt}, {Boccaletti},
  {Bohn}, {Bonnefoy}, {Bonnet}, {Brandner}, {Cantalloube}, {Caselli},
  {Charnay}, {Chauvin}, {Choquet}, {Christiaens}, {Cl{\'e}net}, {Coud{\'e} Du
  Foresto}, {Cridland}, {de Zeeuw}, {Dembet}, {Dexter}, {Drescher}, {Duvert},
  {Eckart}, {Eisenhauer}, {Facchini}, {Gao}, {Garcia}, {Garcia Lopez},
  {Gardner}, {Gendron}, {Genzel}, {Gillessen}, {Girard}, {Haubois},
  {Hei{\ss}el}, {Henning}, {Hinkley}, {Hippler}, {Horrobin}, {Houll{\'e}},
  {Hubert}, {Jim{\'e}nez-Rosales}, {Jocou}, {Kammerer}, {Keppler}, {Kervella},
  {Meyer}, {Kreidberg}, {Lagrange}, {Lapeyr{\`e}re}, {Le Bouquin}, {L{\'e}na},
  {Lutz}, {Maire}, {M{\'e}nard}, {M{\'e}rand}, {Molli{\`e}re}, {Monnier},
  {Mouillet}, {M{\"u}ller}, {Nasedkin}, {Ott}, {Otten}, {Paladini}, {Paumard},
  {Perraut}, {Perrin}, {Pfuhl}, {Pueyo}, {Rameau}, {Rodet},
  {Rodr{\'\i}guez-Coira}, {Rousset}, {Scheithauer}, {Shangguan}, {Shimizu},
  {Stadler}, {Straub}, {Straubmeier}, {Sturm}, {Tacconi}, {van Dishoeck},
  {Vincent}, {von Fellenberg}, {Ward-Duong}, {Widmann}, {Wieprecht},
  {Wiezorrek}, {Woillez}, \& {Gravity Collaboration}}]{Wang2021}
{Wang}, J.~J., {Vigan}, A., {Lacour}, S., {et~al.} 2021,
  \href{http://dx.doi.org/10.3847/1538-3881/abdb2d}{\JournalTitle{\aj}, 161,
  148}

\bibitem[{{Ward-Duong} {et~al.}(2021){Ward-Duong}, {Patience}, {Follette}, {De
  Rosa}, {Rameau}, {Marley}, {Saumon}, {Nielsen}, {Rajan}, {Greenbaum}, {Lee},
  {Wang}, {Czekala}, {Duch{\^e}ne}, {Macintosh}, {Ammons}, {Bailey}, {Barman},
  {Bulger}, {Chen}, {Chilcote}, {Cotten}, {Doyon}, {Esposito}, {Fitzgerald},
  {Gerard}, {Goodsell}, {Graham}, {Hibon}, {Hom}, {Hung}, {Ingraham}, {Kalas},
  {Konopacky}, {Larkin}, {Maire}, {Marchis}, {Marois}, {Metchev},
  {Millar-Blanchaer}, {Oppenheimer}, {Palmer}, {Perrin}, {Poyneer}, {Pueyo},
  {Rantakyr{\"o}}, {Ren}, {Ruffio}, {Savransky}, {Schneider},
  {Sivaramakrishnan}, {Song}, {Soummer}, {Tallis}, {Thomas}, {Wallace},
  {Wiktorowicz}, \& {Wolff}}]{Ward-Duong2021}
{Ward-Duong}, K., {Patience}, J., {Follette}, K., {et~al.} 2021,
  \href{http://dx.doi.org/10.3847/1538-3881/abc263}{\JournalTitle{\aj}, 161, 5}

\end{thebibliography}
\bibliographystyle{yahapj}  % avoids the "----" style, which is nice but makes it more difficult for e.g. NASA ADS to make links



\end{document}